\documentclass[12pt]{article}

\usepackage{amsmath}
\usepackage{amsthm}
\usepackage[margin=1in]{geometry}
\usepackage{thumbpdf,lmodern}
\usepackage{makecell}
\usepackage{tabularx}
\usepackage{graphicx}
\usepackage{multirow}
\usepackage[round]{natbib}
\usepackage{amsfonts}
\usepackage{framed}
\usepackage{hyperref}
\usepackage{authblk}
\usepackage{tablefootnote}
\usepackage{float}
\usepackage{setspace}
\usepackage{caption}
\providecommand{\keywords}[1]
{
  \small	
  \textbf{\textit{Keywords---}} #1
}

\begin{document}

\title{ANPP: the Adapted Normalized Power Prior for Borrowing Information from Multiple Historical Datasets in Clinical Trials}

\author[1]{Yueqi Shen\thanks{angieshen6@gmail.com}}
\author[2]{Matthew A. Psioda}
\author[3]{Luiz M. Carvalho}
\author[1]{Joseph G. Ibrahim}
\affil[1]{Department of Biostatistics, University of North Carolina at Chapel Hill}
\affil[2]{GSK}
\affil[3]{School of Applied Mathematics, Getulio Vargas Foundation}

\maketitle

\abstract{The power prior is a popular class of informative priors for incorporating information from historical data. It involves raising the likelihood for the historical data to a power, which acts as a discounting parameter. When the discounting parameter is modeled as random, the normalized power prior (NPP) is recommended. When there are multiple historical datasets, there has been limited research on how to choose priors for the multiple discounting parameters of the NPP to induce desirable information borrowing behavior. In this work, we address this question by investigating the analytical relationship between the NPP and the Bayesian hierarchical model (BHM), which is a widely used method for synthesizing information from different sources. We develop the adapted normalized power prior (ANPP), which establishes dependence between the dataset-specific discounting parameters of the NPP, leading to inferences that are identical to the BHM. We establish a direct relationship between the prior for the dataset-specific discounting parameters of the ANPP and the prior for the variance parameter of the BHM. Establishing this relationship not only justifies the NPP from the perspective of hierarchical modeling, but also achieves easy prior elicitation for the NPP for the purpose of dynamic borrowing. We examine the borrowing properties of the ANPP through simulations, and apply it to a case study for a pediatric lupus trial.}


\section{Introduction}

Borrowing information from historical trials has become an increasingly common strategy in modern drug development, particularly in therapeutic areas such as pediatric trials and rare diseases. The appropriate use of historical data in clinical trials increases the effective sample size, which can potentially significantly improve the efficiency of clinical trials. Utilizing Bayesian methods to incorporate historical information through informative priors is a natural approach. In the case of pediatric trials, for example, FDA’s ICH Guideline on pediatric extrapolation \citep{fda_e11a} includes Bayesian strategies as potential study designs for borrowing information from adult studies. 

The power prior \citep{chen_2000} is a popular class of informative priors that allow the incorporation of historical data through a discounted likelihood.
It is constructed by raising the historical data likelihood to a power $a_0$, where $0 \le a_0 \le 1$.
The discounting parameter $a_0$ can be fixed or modeled as random.
When it is modeled as random and estimated jointly with other parameters of interest (denoted by $\theta$), the normalized power prior (NPP) \citep{duan_2006} is recommended as normalization is critical to enabling the prior to factor into a conditional distribution of $\theta$ given $a_0$ and a marginal distribution of $a_0$ \citep{neuens_2009}. Theoretical properties of the NPP have been extensively studied in the works of \cite{Carvalho_Ibrahim_2021}, \cite{YE202229}, \cite{shen_2023}, \cite{pawel_2023} and \cite{Han_Ye_Wang_2022}.
For example, \cite{Carvalho_Ibrahim_2021} show that the NPP is always proper when the initial prior is proper, and that, viewed as a function of the discounting parameter, the normalizing constant is a smooth and strictly convex function. \cite{YE202229} prove that if the prior on $a_0$ is non-decreasing and if the difference between the sufficient statistics of the historical and current data is negligible from a practical standpoint, the marginal posterior mode of $a_0$ is close to one. \cite{shen_2023} provide a formal analysis of the asymptotic behavior of the marginal posterior for $a_0$ for generalized linear models (GLMs). They show that the marginal posterior for the discounting parameter converges
to a point mass at zero if there is any discrepancy between the historical and current data. They also show that the marginal posterior for $a_0$ does not converge to a point mass at one when the datasets are fully compatible, and yet, for an i.i.d. normal model and finite sample size, the marginal posterior for $a_0$ always has most mass around one when the datasets are fully compatible. 

When there is assumed to be a single historical dataset, several methods have been proposed which provide guidance on choosing the discounting parameter $a_0$ for the power prior, or the prior for $a_0$ for the NPP.
Many quasi-Bayesian or empirical Bayes-type approaches have been developed to adaptively determine $a_0$ in the power prior based on the degree of congruence between the historical and current data (\cite{Gravestock_Held_2017,Gravestock_Held_2019}; \cite{Liu_2018}; \cite{Bennett_2021}; \cite{Pan_Yuan_Xia_2017}; \cite{psioda_ped}).
\cite{shen_2023} provide a fully Bayesian approach for eliciting optimal beta priors on $a_0$ for the NPP. \cite{chen_2006} provide insights into eliciting a guide value for $a_0$ in the power prior using the Bayesian hierarchical models (BHM), one of the most common methods for synthesizing information from different sources \citep{gelman}, by establishing an analytic relationship between $a_0$ and the variance parameter of the BHM.
This analytic relationship was also established for generalized linear models and extended to the case with multiple historical datasets.
\cite{pawel_test} generalize this connection to the NPP, and show that for normal models with a single historical dataset, inferences with the NPP using a beta prior on $a_0$ align with inferences with the BHM using a generalized beta prior on the relative heterogeneity variance. This connection between the NPP and BHM has not been extended to the case with multiple historical datasets, however. The main challenge is that the $a_0$ in the NPP becomes multi-dimensional but the variance parameter in the BHM is one-dimensional.   

So far in the literature, there has been limited research that investigates the NPP in the context of multiple historical datasets. \cite{han_2025} introduce the ordered NPP when a natural a priori ordering arises among the historical trials based on factors such as baseline characteristics and the year the trial was conducted. The ordered NPP imposes a targeted order restriction on the discounting parameters. More generally, \cite{Banbeta_2019} develop the dependent and robust dependent NPP which allows dependent discounting parameters for multiple historical datasets by placing a hierarchical beta prior on them. While this model is useful and intuitive, tuning the hyperparameters of the beta prior is not a straightforward task. The question of how to choose an optimal prior on the  discounting parameters of the NPP in order to synthesize information from a number of historical datasets and appropriately borrow from each of them remains unanswered. 


In this paper, our goal is to extend the work in \cite{shen_2023} to study the choice of prior on discounting parameters in the NPP for multiple hisotrical datasets. We take a different approach from \cite{Banbeta_2019}, and instead focus on establishing a connection between the NPP and the BHM for multiple historical datasets, extending the work of \cite{chen_2006} and \cite{pawel_test}. To this end, we propose a novel model, the adapted normalized power prior (ANPP), that establishes dependence between the multiple discounting parameters for the multiple historical datasets, resulting in inference for the parameter of interest that is identical to inference based on the BHM. The ANPP allows for dependent discounting of the historical datasets without the need to specify a prior for each discounting parameter, since each discounting parameter is shown to be a deterministic function of a single global discounting parameter. In particular, we establish a direct relationship between the prior for the discounting parameters of the ANPP and the prior for the variance parameter of the BHM that leads to equivalent posterior inference for the two models. Establishing this relationship for multiple historical datasets is an entirely novel development. Establishing this relationship not only justifies the ANPP from the perspective of hierarchical modeling, but also provides insight on how one might elicit the prior for the discounting parameters through the lens of hierarchical modeling.
If one starts with an appropriate prior on the variance parameter in a BHM, a prior on the discounting parameters in an ANPP is induced, providing a semi-automatic benchmark prior for the discounting parameters in the ANPP. In addition, when adjusting the amount of information to borrow, the calibration of a prior on a parameter on the scale of $0 < a_0 < 1$ is more intuitive than that of a variance parameter. Since we prove that there are prior choices that lead to equivalent posterior inference on $\theta$, using the ANPP has the advantage of easier prior calibration for the purpose of dynamic borrowing compared to the BHM. We extensively study the borrowing properties of the ANPP through simulations, and present strategies on inducing priors on the discounting parameters based on the BHM. We also demonstrate how our theoretical results can be applied more generally for non-normal data using asymptotic approximations through a case study.

The rest of the paper proceeds as follows.
In Section 2, we define the ANPP for multiple historical datasets and show the sufficient conditions under which it will produce equivalent posterior inference for the parameter of interest compared to the BHM for the normal model.
In Section 3, we explore the borrowing properties of the ANPP through extensive simulations and compare its performance to the NPP with independent priors on the discounting parameters. We also visually present the induced priors on $a_0$ based on common choices of prior for the variance parameter in the BHM, and vice versa. In Section 4, we apply the ANPP to real data from a pediatric lupus trial where the prior is formulated to borrow information from two adult trials. In Section 5, we close the paper with some discussion and thoughts on future research. 

\section{Method}
\label{sec:connection}

\subsection{Theoretical Background for the Single Historical Dataset Case}

We first lay out the mathematical relationship between the NPP and the BHM for a single historical dataset, as this relationship motivates our method for adapting the NPP for multiple historical datasets.

We first consider \textit{i.i.d.} normal data with a single historical dataset.
Let $y = (y_1, \dots, y_n)$ denote the current data with sample size $n$. The current data model is $$y_i \sim N(\theta, \sigma^2),$$ where $i = 1, \dots, n$, and $\sigma^2$ is fixed and known. Let $y_0 = (y_{01}, \dots, y_{0n_0})$ denote the historical data with sample size $n_0$.
The historical data model is $$y_{0i} \sim N(\theta_0, \sigma^2_0),$$ where $i = 1, \dots, n_0$, and $\sigma^2_0$ is fixed and known.

The Bayesian hierarchical model (BHM) commonly uses priors of the form
$$\theta|\mu,v \sim N(\mu, v), \qquad \theta_0|\mu,v \sim N(\mu, v),$$
and hyperpriors
$$\mu \sim N(\alpha, \nu^2), \qquad v \sim IG(c, d),$$
where $\alpha$, $\nu$, $c$ and $d$ are fixed hyperparameters. Note that our theoretical development does not require the prior on $v$ to be the inverse Gamma distribution. For example, a half-normal prior as advocated by \cite{gelman} could easily be used. 

The joint posterior for $\theta$, $\theta_0$, $\mu$, and $v$ based on the BHM is given by
\begin{tiny}
\begin{align*}
\pi(\theta,\theta_0,\mu,v|y, y_0) \propto \frac{1}{v}\exp\left\{-\frac{n(\bar{y}-\theta)^2}{2\sigma^2} - \frac{n_0(\bar{y}_0-\theta_0)^2}{2\sigma_0^2}-\frac{(\theta-\mu)^2+(\theta_0-\mu)^2}{2v}-\frac{(\mu-\alpha)^2}{2\nu^2}\right\}\pi(v),
\end{align*}
\end{tiny}
and the marginal posterior for the main effect $\theta$ is given by
\begin{tiny}
\begin{align}\label{post_bhm}
\pi(\theta|y, y_0) \propto &\int\int\int&\frac{1}{v}\exp\left\{-\frac{n(\bar{y}-\theta)^2}{2\sigma^2} - \frac{n_0(\bar{y}_0-\theta_0)^2}{2\sigma_0^2}-\frac{(\theta-\mu)^2+(\theta_0-\mu)^2}{2v}-\frac{(\mu-\alpha)^2}{2\nu^2}\right\}\pi(v)d\theta_0 d\mu dv.
\end{align}
\end{tiny}
The power prior \citep{chen_2000} is formulated as 
\begin{align*}
\pi(\theta|D_0, a_0) \propto L(\theta|D_0)^{a_0}\pi_0(\theta),
\end{align*}
where $0 \le a_0 \le 1$ is the discounting parameter which discounts the historical data likelihood, and $\pi_0(\theta)$ is the initial prior for $\theta$. When $a_0$ is modeled as random, the normalized power prior \citep{duan_2006} (NPP) is recommended. The NPP is formulated as
$$\pi(\theta,a_0|y_0) \propto \frac{L(\theta|y_0)^{a_{0}}\pi_0(\theta)}{c(a_0)}\pi(a_0),
$$
where $\pi(a_0)$ is the prior for $a_0$ and $c(a_0)=\int L(\theta|y_0)^{a_{0}}\pi_0(\theta)d\theta$ is the normalizing constant which is a function of $a_0$.
For the normal model specified above with the initial prior $\pi_0(\theta) \propto 1$, the NPP becomes
$$\pi(\theta,a_0|y_0) \propto \frac{\left[\sigma_0^{-n_{0}} \exp \left\{-\frac{1}{2 \sigma_{0}^{2}} \sum_{i=1}^{n_{0}}\left(y_{0i}-\theta\right)^{2}\right\}\right]^{a_{0}} }{c(a_0)}\pi(a_0),$$
where $c(a_{0})=\int \left[\sigma_{0}^{-n_{0}} \exp \left\{-\frac{1}{2 \sigma_{0}^{2}} \sum_{i=1}^{n_{0}}\left(y_{0i}-\theta\right)^{2}\right\}\right]^{a_{0}} d\theta$.
Then the marginal posterior for $\theta$ based on the NPP is
\begin{align}\label{post_npp}
\pi(\theta|y, y_0) \propto \int\exp \left\{-\frac{1}{2 \sigma^{2}} \sum_{i=1}^{n}\left(y_{i}-\theta\right)^{2}\right\}\frac{\left[\sigma_0^{-n_{0}} \exp \left\{-\frac{1}{2 \sigma_{0}^{2}} \sum_{i=1}^{n_{0}}\left(y_{0i}-\theta\right)^{2}\right\}\right]^{a_{0}} }{c(a_0)}\pi(a_0)da_0.
\end{align}

When the parameter $v$ in the BHM and the parameter $a_0$ in the NPP are fixed, Theorem 2.2 in \cite{chen_2006} states that the posterior for $\theta$ based on the BHM and the posterior for $\theta$ based on the power prior match if and only if the prior expectation $E_{\text{BHM}}[\mu] = \alpha = 0$, $\nu^2 \rightarrow \infty$, and
$$a_0=\frac{1}{\frac{2v n_0}{\sigma_0^2}+1}.$$
This result connects the BHM and the power prior when $a_0$ is fixed. 

In equation (20) in \cite{pawel_test}, this theorem is extended to scenarios where $v$ is modeled as random for the BHM and $a_0$ is modeled as random for the NPP. Specifically, it is shown that there are in fact choices for (hyper)priors $\pi(v)$ in the BHM and $\pi(a_0)$ in the NPP that lead to equivalent marginal posteriors for $\theta$. Let $\pi_\text{NPP}(\theta|a_0)$ denote the posterior for $\theta$ conditional on $a_0$ using the NPP and $\pi_\text{BHM}(\theta|v)$ denote the posterior for $\theta$ conditional on $v$ using the BHM.
The goal is to find a transformation from $a_0$ to $v$ such that the marginals of $\theta$ based on the two models are equivalent, i.e.,
$$\int\pi_{\text{NPP}}(\theta|a_0)\pi(a_0)da_0=\int\pi_{\text{BHM}}(\theta|v)\pi(v)dv.$$
That is, the goal is to find a function $a_0=f(v)$ such that
$$\int\pi_{\text{NPP}}(\theta|f(v))\pi(f(v))\left\vert\frac{df(v)}{dv}\right\vert dv=\int\pi_{\text{BHM}}(\theta|v)\pi(v)dv.$$

\newtheorem{thm}{Theorem}[section]
\newtheorem{lem}[thm]{Lemma}
\newtheorem{cor}{Corollary}[section]

The resulting $f(v)$ and conditions for priors for $a_0$ and $v$ are presented in Theorem \ref{single} below, similar to equation (20) in \cite{pawel_test}. Here, we present an alternative proof to the proof used in \cite{pawel_test}, as our results are derived independently. 

\begin{thm}\label{single}
The marginal posterior for $\theta$ based on the BHM given in \eqref{post_bhm} and the marginal posterior for $\theta$ based on the NPP given in \eqref{post_npp} are identical if and only if $$a_0=f(v)=\frac{1}{\frac{2vn_0}{\sigma_0^2}+1} $$ and if $$\frac{2n_0}{\sigma_0^2}\left(\frac{2v n_0}{\sigma_0^2}+1\right)^{-2}\pi(f(v))=\pi(v).$$
\end{thm}
\begin{proof}
See Appendix \ref{appen:th1}. 
\end{proof}

Theorem \ref{single} gives us a direct relationship between the prior for $a_0$ in the NPP and the prior for the variance parameter of the BHM. Note that the transformation from $a_0$ to $v$ used here is the same as in Theorem 2.2 in \cite{chen_2006}.
In Appendix \ref{appen:fig}, we provide Figure \ref{test_single} that illustrates empirically through sampling that the formulas in Theorem \ref{single} lead to equivalent posteriors for $\theta$ based on the BHM and the NPP.

\subsection{The Adapted Normalized Power Prior (ANPP) for Multiple Historical Datasets}
We now proceed to develop the adapted normalized power prior (ANPP), which results in identical posterior inference to the posterior inference using the BHM with certain (hyper)prior choices in the presence of multiple historical datasets. 

Consider \textit{i.i.d.} normal data with historical datasets $k=1,\dots, K$.  Let $y = (y_1, \dots, y_n)$ denote the current data with sample size $n$.
The model for the current data is
$$y_i \sim N(\theta, \sigma^2),$$
where $i = 1, \dots, n$, and $\sigma^2$ is fixed and known.
Let $y_0 = (y_{01}, \dots, y_{0K})$ denote the $K$ historical datasets each with sample size $n_{01},\dots,n_{0K}$, where $y_{0k}=(y_{0k1},\dots,y_{0kn_{0k}})$ for $k=1,\dots, K$.
The historical data model is $$y_{0ki} \sim N(\theta_{0k}, \sigma^2_{0k}),$$ where $i = 1, \dots, n_{0k}$, and $\sigma^2_{0k}$'s are fixed and known.
Let $\theta_0=(\theta_{01},\dots,\theta_{0K})$.

The Bayesian hierarchical model (BHM) commonly uses priors that take the form $$\theta|\mu,v \sim N(\mu, v), \qquad \theta_{0k}|\mu,v \sim N(\mu, v),$$ and $$\mu \propto 1, \qquad v \sim IG(c, d),$$ where $c$ and $d$ are fixed hyperparameters. We use $\pi(\mu) \propto 1$ here because the result in Theorem \ref{single} can be obtained alternatively by using a uniform improper prior for $\mu$ at the outset (see Corollary 2.2. in \cite{chen_2006}).
Then the joint posterior for $\theta$, $\theta_0$, $\mu$, and $v$ based on the BHM is given by
\begin{tiny}
\begin{align*}
\pi(\theta, \theta_0, \mu, v|y, y_0) \propto & \hspace{0.2cm}\pi(v)v^{-(K+1)/2}\\
&\exp\left\{\frac{-n(\bar{y}-\theta)^2}{2\sigma^2}-\frac{1}{2}\sum_{k=1}^{K}\frac{n_{0k}(\bar{y}_{0k}-\theta_{0k})^2}{\sigma^2_{0k}}-\frac{(\theta-\mu)^2}{2v}-\frac{1}{2v}\sum_{k=1}^{K}(\theta_{0k}-\mu)^2\right\},
\end{align*}
\end{tiny}
and the marginal posterior for $\theta$ is given by
\begin{tiny}
\begin{align}\label{mul_bhm}
\pi(\theta|y, y_0) \propto \int\int\int \pi(v)v^{-(K+1)/2}
\exp\left\{\frac{-n(\bar{y}-\theta)^2}{2\sigma^2}-\frac{1}{2}\sum_{k=1}^{K}\frac{n_{0k}(\bar{y}_{0k}-\theta_{0k})^2}{\sigma^2_{0k}}-\frac{(\theta-\mu)^2}{2v}-\frac{1}{2v}\sum_{k=1}^{K}(\theta_{0k}-\mu)^2\right\} d\theta_0 d\mu dv.
\end{align}
\end{tiny}

When there are multiple historical datasets, we assume $a_0=(a_{01},\dots,a_{0K})$ are the discounting parameters with $a_{0k}$ corresponding to historical dataset $k$.
When the $a_{0k}$'s are given independent priors, the normalized power prior (henceforth abbreviated iNPP) for $\theta$ and $a_0$ is given by
$$\pi_{iNPP}(\theta,a_0|y_0) \propto \frac{\prod_{k=1}^{K}\left[\sigma_{0 k}^{-n_{0k}} \exp \left\{-\frac{1}{2 \sigma_{0 k}^{2}} \sum_{i=1}^{n_{0 k}}\left(y_{0ki}-\theta\right)^{2}\right\}\right]^{a_{0 k}} }{c(a_0)}\prod_{k=1}^{K}\pi(a_{0k}),$$
with
$$c(a_{0})=\int \prod_{k=1}^{K}\left[\sigma_{0 k}^{-n_{0k}} \exp \left\{-\frac{1}{2 \sigma_{0 k}^{2}} \sum_{i=1}^{n_{0 k}}\left(y_{0ki}-\theta\right)^{2}\right\}\right]^{a_{0 k}} d\theta.$$
In order to establish a correspondence between the NPP and the BHM for multiple historical datasets, we must assume some relationship between the discounting parameters in the NPP. 

We define the adapted normalized power prior (ANPP) as follows:
$$\pi_{ANPP}(\theta,a_0|y_0) \propto \frac{\prod_{k=1}^{K}\left[\sigma_{0k}^{-n_{0k}} \exp \left\{-\frac{1}{2 \sigma_{0 k}^{2}} \sum_{i=1}^{n_{0 k}}\left(y_{0ki}-\theta\right)^{2}\right\}\right]^{h_k(a_0)}}{c(a_0)}\pi(a_0),$$
where $0 < h_k(a_0) < 1$ is a transformation of $a_0$ which discounts the $k^{th}$ historical dataset, and $$c(a_{0})=\int\prod_{k=1}^{K}\left[\sigma_{0 k}^{-n_{0k}} \exp \left\{-\frac{1}{2 \sigma_{0 k}^{2}} \sum_{i=1}^{n_{0 k}}\left(y_{0ki}-\theta\right)^{2}\right\}\right]^{h_k(a_0)} d\theta.$$
Then the marginal posterior for $\theta$ with the ANPP is
\begin{tiny}
\begin{align}\label{mul_npp}
\pi_{ANPP}(\theta|y, y_0) \propto \int\exp\left\{-\frac{1}{2\sigma^2}\sum_{i=1}^n(y_i-\theta)^2\right\}\frac{\prod_{k=1}^{K}\left[\sigma_{0k}^{-n_{0k}} \exp \left\{-\frac{1}{2 \sigma_{0 k}^{2}} \sum_{i=1}^{n_{0 k}}\left(y_{0ki}-\theta\right)^{2}\right\}\right]^{h_k(a_0)}}{c(a_0)}\pi(a_0)da_0.
\end{align}
\end{tiny}
Our goal is to characterize $h_k(a_0)$ and the transformation $a_{0}=f(v)$ such that
$$\int\pi_{\text{ANPP}}(\theta|f(v))\pi(f(v))\left\vert\frac{df(v)}{dv}\right\vert dv=\int\pi_{\text{BHM}}(\theta|v)\pi(v)dv.$$ 

\begin{thm}\label{multiple}
The marginal posterior for $\theta$ based on the BHM given in \eqref{mul_bhm} and the marginal posterior for $\theta$ based on the ANPP given in \eqref{mul_npp} are identical if $$a_0=f(v)=\left[1+\sum_{k=1}^{K}\frac{\frac{n_{0k}}{\sigma^2_{0k}}}{\frac{n_{0k}}{\sigma^2_{0k}}+\frac{1}{v}}\right]^{-1},$$ $$h_k(a_0)=\left(1+\frac{n_{0k}v}{\sigma^2_{0k}}\right)^{-1}a_0,$$ and $$Q(f(v)) \cdot \left\vert\frac{df(v)}{dv}\right\vert \pi(f(v)) = R(v) \cdot \pi(v),$$ where 
$$Q(f(v))=\exp\left\{-\frac{f(v)}{2C}\left(\frac{1}{v}\sum_{k=1}^{K}N_kY_k\right)^2\right\}\left[f(v)C\right]^{\frac{1}{2}}$$ and 
$$R(v)=v^{-(K+1)/2}\prod_{k=1}^{K}N_k^{\frac{1}{2}}A^{-\frac{1}{2}}\exp\left\{\frac{\left(\sum_{k=1}^{K}Y_kN_k\right)^2}{2v^2A}\right\}\exp\left\{\frac{1}{2}\sum_{k=1}^{K}Y_k^2N_k\right\},$$
with $N_k=\left(\frac{n_{0k}}{\sigma^2_{0k}}+\frac{1}{v}\right)^{-1}$, $C=\sum_{k=1}^{K}\frac{n_{0k}N_k}{v\sigma^2_{0k}}$, $Y_k=\frac{n_{0k}\bar{y}_{0k}}{\sigma^2_{0k}}$ and $A=\frac{1+K}{v} - \frac{\sum_{k=1}^{K}N_k}{v^2}$.
The Jacobian $\left\vert\frac{df(v)}{dv}\right\vert$ is given in the proof. 
\end{thm}
\begin{proof}
See Appendix \ref{appen:th2}.
\end{proof}

The term $h_k(a_0)$ in the ANPP is a function of the global discounting parameter $a_0$ and the variance and sample size of the $k^{th}$ historical dataset. The ANPP allows for dependent discounting of the historical datasets without the need to specify a prior for each discounting parameter, since each discounting parameter is shown to be a deterministic function of the global $a_0$. 
Theorem \ref{multiple} gives a direct relationship between the prior for $a_0$ in the ANPP and the prior for the variance parameter of the BHM. 
It enables the elicitation of priors for the multiple discounting parameters through the lens of
hierarchical modeling. If one starts with an appropriate prior on the variance parameter in a BHM, a prior on the discounting parameters in an ANPP is induced, providing a semi-automatic benchmark prior for the discounting parameters in the ANPP. To validate the results in Theorem \ref{multiple} empirically, we present Figure \ref{test_multiple} in Appendix \ref{appen:fig} which illustrates through sampling, implemented with RStan \citep{rstan}, that the formulas in Theorem \ref{multiple} lead to equivalent posteriors for $\theta$ based on the BHM and ANPP.

\section{Simulations}

\subsection{Comparison of ANPP and iNPP}

In this section, we explore the behavior of the ANPP and compare it to the NPP with independent $a_0$'s (iNPP) for various generated data scenarios with multiple historical datasets from which to borrow information.
In the following set of simulations, we assume the current data $y_1, \dots, y_n$ are i.i.d. observations from a $N(\theta,\sigma^2)$ distribution.
We also assume there are two historical datasets, where $y_{0k1}, \dots, y_{0kn_{0k}}$ are i.i.d. observations from a $N(\theta, \sigma^2_{0k})$ where $k=1,2$.
Let $\bar{y}$ denote the mean of the current data, and $\bar{y}_{01}$ and $\bar{y}_{02}$ denote the mean of the first and second historical datasets, respectively.
Let $a_{01}$ and $a_{02}$ denote the discounting parameters for the first and second historical datasets, respectively.
We assume independent uniform priors for $a_{01}$ and $a_{02}$ for the iNPP.
For the ANPP, $a_{01}=h_1(a_0)$ and $a_{02}=h_2(a_0)$, and the prior on $v$ is chosen so that the conditions in Theorem \ref{multiple} are satisfied where the prior on $a_0$ is the uniform distribution.
We choose $\sigma^2=\sigma^2_{01}=\sigma^2_{02}=1$, $\bar{y}=0$ and $n=30$.
We vary the means and the sample sizes of the historical datasets to investigate the performance of the ANPP and iNPP under various degrees of compatibility between the current and historical datasets.
Specifically, through the marginal posteriors for $a_{01}$, $a_{02}$ and $\theta$, we examine how much historical information is borrowed and its impact on the inference of the main effect $\theta$ using the ANPP and iNPP.
The ANPP is implemented with RStan \citep{rstan} and the iNPP is implemented using the R package BayesPPD \citep{shen_RJ, bayesppd}. 

For the ANPP, since $a_{0k}$ is a function of only $\sigma^2_{0k}/n_{0k}$ and $v$, and we fix $\sigma^2_{01} = \sigma^2_{02} = 1$ for these simulations, the posterior densities of $a_{01}$ and $a_{02}$ will only differ if $n_{01}$ and $n_{02}$ differ.
In addition, since $a_{0k}$ monotonically increases with $\sigma^2_{0k}/n_{0k}$, given equal variances for the historical datasets, the posterior density of $a_{0k}$ is more concentrated near zero, i.e., the historical dataset is discounted more, for larger sample sizes.
For the BHM, it is well known that groups that have smaller sample size experience greater shrinkage toward the grand mean \citep{berry}, so for the NPP, the smaller historical dataset has to be weighted comparatively more for the NPP to match the BHM. 

\begin{figure}[H]
\centering
\includegraphics[width=0.9\textwidth]{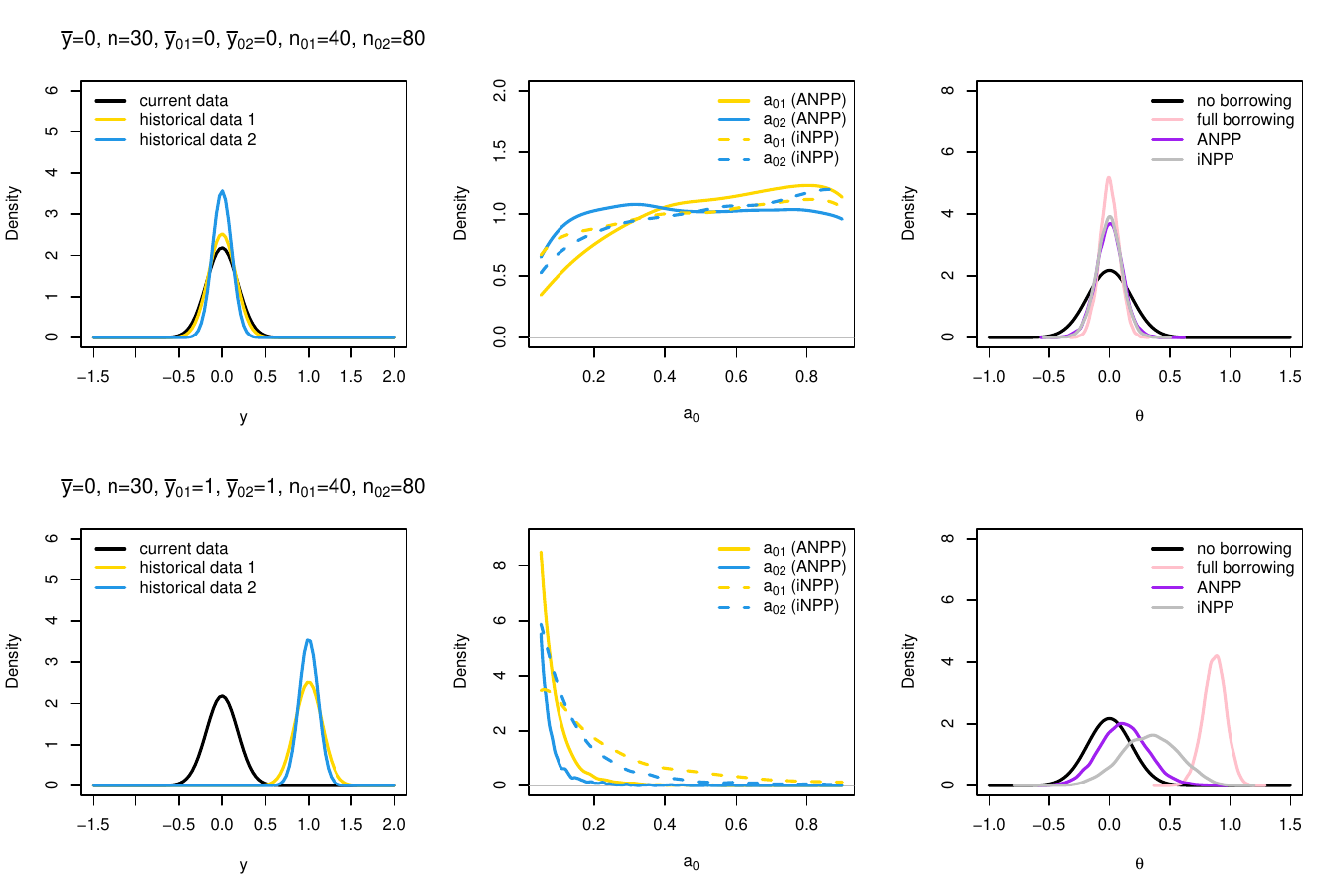}
\captionsetup{width=0.8\linewidth}
\caption{Marginal posteriors for $a_{01}$, $a_{02}$ and $\theta$ using the ANPP and the iNPP for simulated i.i.d. normal data. In the top row, the three datasets are fully compatible.
In the bottom row, the two historical datasets are compatible but they are both incompatible with the current dataset.
The first column includes plots of the densities of the current data (black line) and two historical datasets (yellow and blue lines).
The second column includes plots of the posterior densities of $a_{01}$ and $a_{02}$ using the ANPP (yellow and blue solid lines) and the iNPP (yellow and blue dashed lines).
The third column includes plots of the posterior densities of $\theta$ using four different priors, the ANPP (purple), the iNPP (grey), the power prior with $a_0=0$ (black) and the power prior with $a_0=1$ (pink). }
\label{multiple_sim1}
\end{figure}

\begin{figure}[H]
\begin{center}
\includegraphics[width=0.9\textwidth]{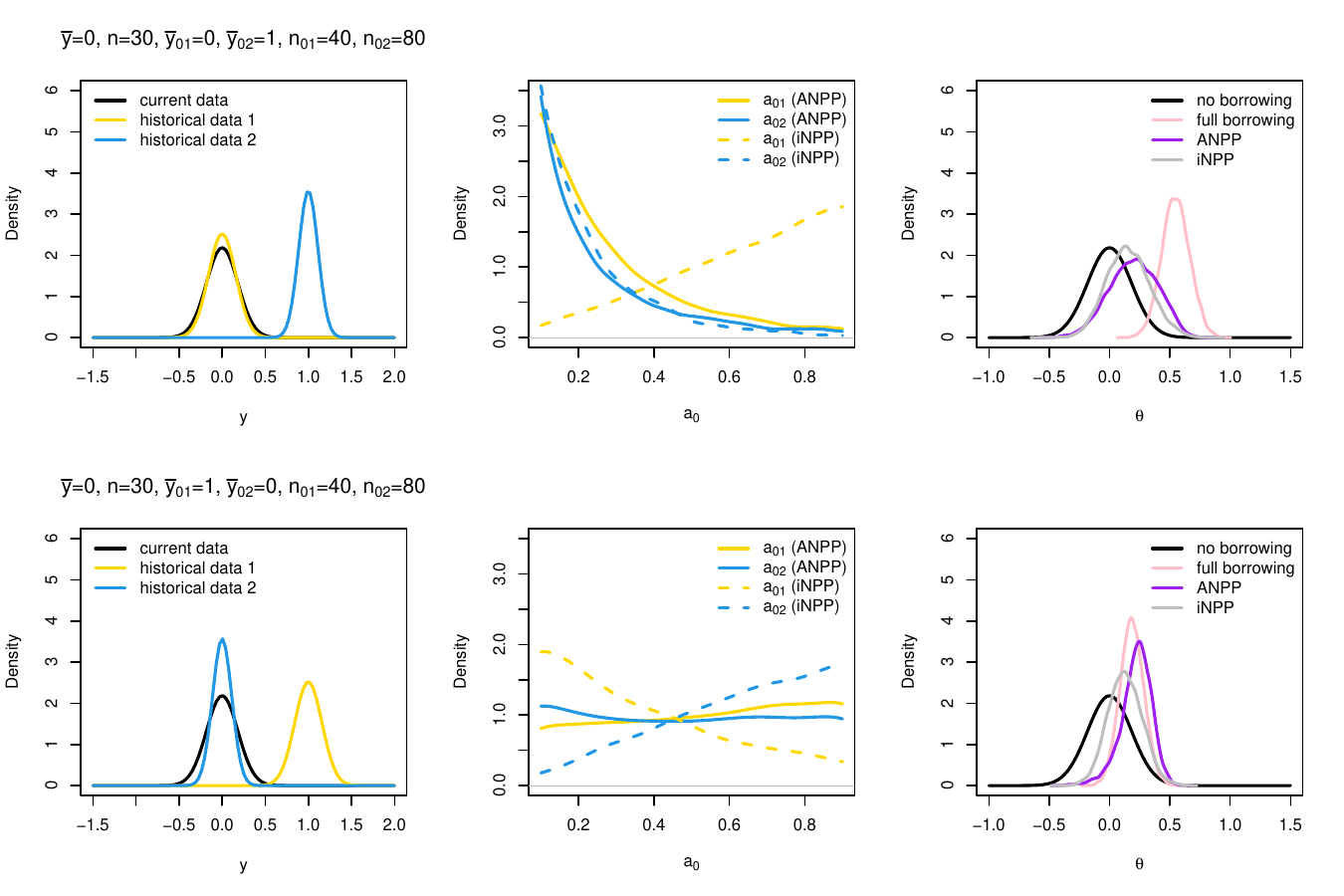}
\captionsetup{width=0.8\linewidth}
\caption{Marginal posteriors for $a_{01}$, $a_{02}$ and $\theta$ using the ANPP and the iNPP for simulated i.i.d. normal data. In these scenarios, one of the historical datasets is compatible with the current data while the other is not. The first column includes plots of the densities of the current data (black line) and two historical datasets (yellow and blue lines).
The second column includes plots of the posterior densities of $a_{01}$ and $a_{02}$ using the ANPP (yellow and blue solid lines) and the iNPP (yellow and blue dashed lines).
The third column includes plots of the posterior densities of $\theta$ using four different priors, the ANPP (purple), the iNPP (grey), the power prior with $a_0=0$ (black) and the power prior with $a_0=1$ (pink). }
\label{multiple_sim2}
\end{center}
\end{figure}

In Figures \ref{multiple_sim1} and \ref{multiple_sim2}, the first column includes density plots of the current data (black line) and two historical datasets (yellow and blue lines).
The second column includes plots of the posterior densities of $a_{01}$ and $a_{02}$ using the ANPP (yellow and blue solid lines) and the iNPP (yellow and blue dashed lines).
The third column includes plots of the posterior densities of $\theta$ using four different priors, the ANPP (purple), the iNPP (grey), the power prior with $a_0=0$ (black) and the power prior with $a_0=1$ (pink).
In the top row of Figure \ref{multiple_sim1}, the current and historical datasets have the same mean, and the posteriors for $a_{01}$ and $a_{02}$ are similar using the ANPP and iNPP, and they both have modes at one. We also observe that the iNPP borrows more from the larger dataset compared to the ANPP.
The posteriors for $\theta$ using the two priors are both centered at zero, yielding smaller variance than the analysis with the power prior that takes $a_0=0$.
In the bottom row, the two historical datasets have the same mean but that mean differs substantially from the current data mean.
In this case, the posteriors for $a_{01}$ and $a_{02}$ using the ANPP are concentrated near zero, while the iNPP results in significantly less discounting.
As expected, compared to the posterior for $\theta$ using the iNPP (grey), the posterior for $\theta$ using the ANPP (purple) is much closer to that using only the current data (black).

In Figure \ref{multiple_sim2}, we assume one of the historical datasets is compatible with the current data while the other is not.
In the top row, the historical dataset with the greater sample size is incompatible with the current data, while in the bottom row, the historical dataset with the greater sample size is compatible with the current data.
We observe that, for both cases, the iNPP leads to more borrowing from the compatible historical dataset and less borrowing from the incompatible one in comparison to the ANPP.
The ANPP discounts in accordance with the overall amount of conflict between the three datasets, and discounts more for more incompatible datasets (e.g., the top row).
In the top row, the posterior for $\theta$ using the iNPP is slightly closer to that using only the current data compared to the ANPP, due to less overall borrowing. In the bottom row, we can see that the posterior mean of $\theta$ is larger for the iNPP than the full borrowing case (power prior with $a_0=1$); this is because the first historical dataset has the same mean as the current dataset and full borrowing will lead to a mean closer to the current data mean.
Moreover, the posterior for $\theta$ with iNPP is closer to that using only current data compared to the ANPP; this is because the iNPP leads to more borrowing from the compatible historical dataset and less from the incompatible historical dataset when compared to the ANPP.

In summary, the ANPP and iNPP behave similarly when the historical and current datasets are compatible. 
The ANPP is more sensitive to conflicts in the data than the iNPP. The ANPP tends to borrow more from the smaller dataset, which is a well-known property of the BHM as well \citep{berry}.
When one of the historical datasets is compatible with the current data while the other is not, the iNPP borrows more from the compatible historical dataset compared to the ANPP which effectively bases discounting on the overall heterogeneity across all datasets.
We include additional simulations when the historical datasets are partially compatible with the current dataset in Figure \ref{multiple_sim3} in Appendix \ref{appen:fig}. 

\subsection{Visual Presentation of Prior Relationships}

\begin{figure}[H]
\begin{center}
\includegraphics[width=0.9\textwidth]{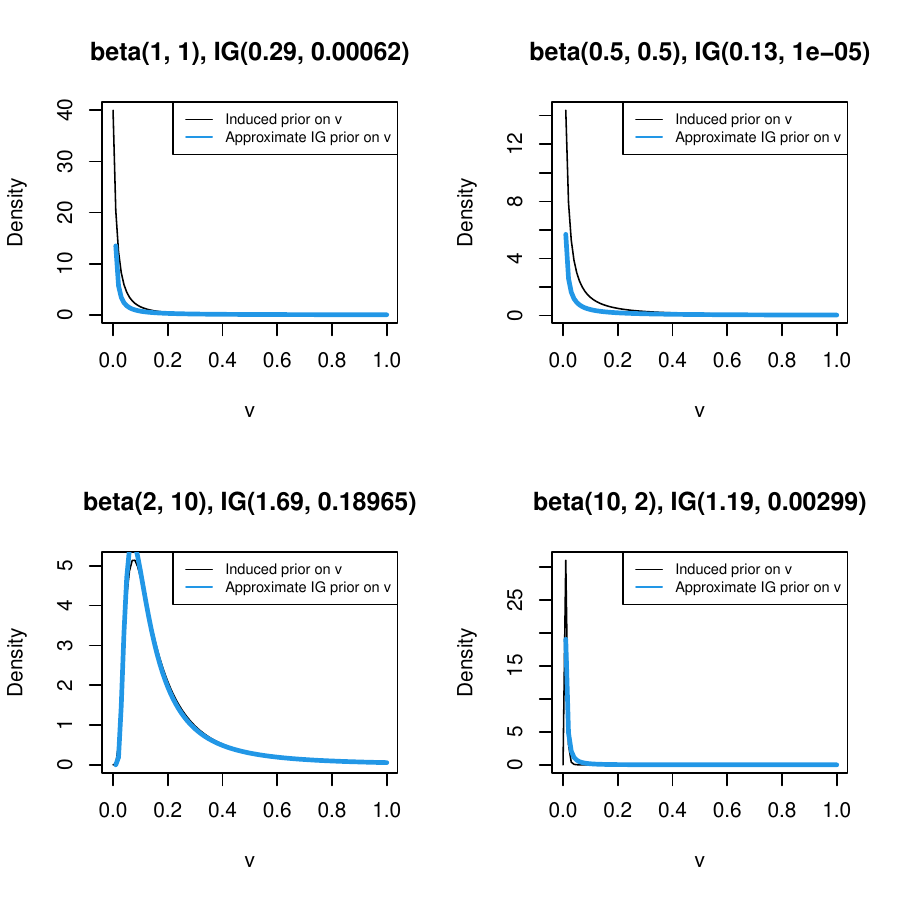}
\captionsetup{width=0.8\linewidth}
\caption{We choose four beta distributions as the prior for $a_0$ in the NPP, and find the best approximating inverse gamma distribution prior for $v$ in a BHM.
The black line represents the induced prior on $v$ using Theorem \ref{single}, and the blue line represents the inverse gamma distribution that best  approximates the induced prior.}
\label{single_beta}
\end{center}
\end{figure}

\begin{figure}[H]
\begin{center}
\includegraphics[width=0.9\textwidth]{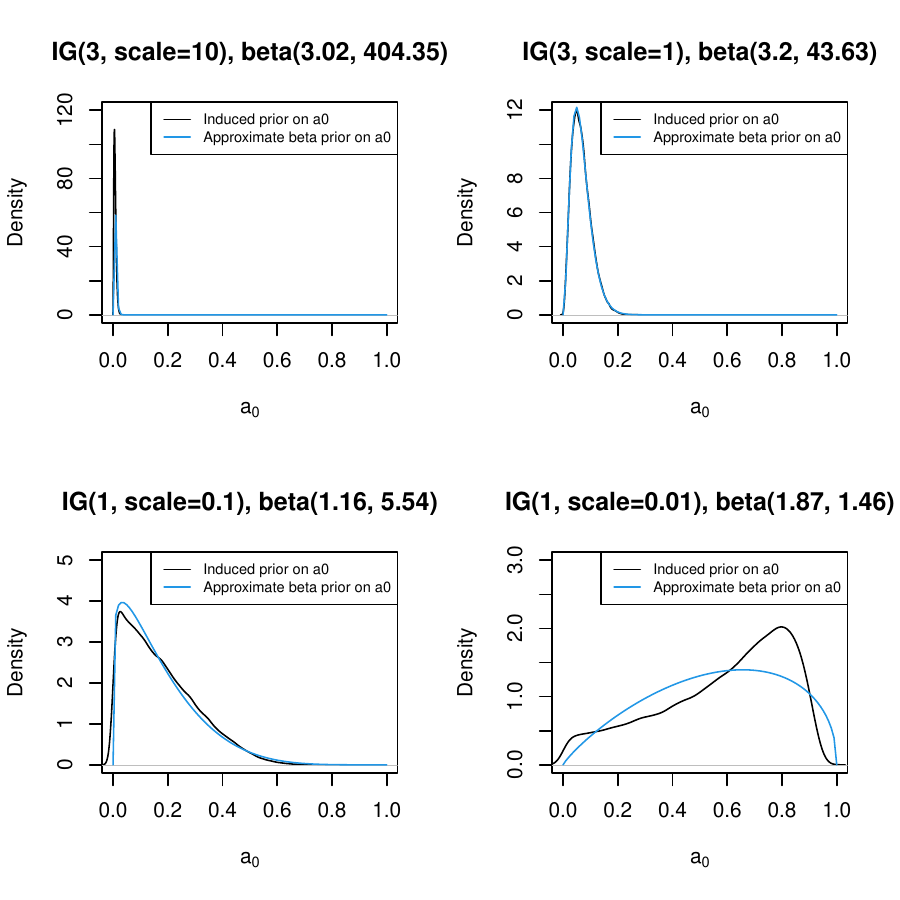}
\captionsetup{width=0.8\linewidth}
\caption{We choose four inverse gamma distributions as the prior for $v$ in the BHM, and find the best approximating beta prior for $a_0$ in a NPP.
We draw samples of $v$ from the inverse gamma distributions and transform them to samples of $a_0$ using the formula in Theorem \ref{single}.
We use beta regression to solve for the beta distribution that best fits the $a_0$ samples.
The black line represents the induced prior on $a_0$ using Theorem \ref{single}, and the blue line represents the beta distribution that best approximates the induced prior.}
\label{single_IG}
\end{center}
\end{figure}

In this section, we provide some visual intuition on the connection between the prior on $a_0$ for the NPP and the prior on $v$ for the BHM when there is a single historical dataset. Since the formulas in Theorems \ref{single} and \ref{multiple} may not be conducive to an intuitive understanding of the relationship between the two priors, we provide Figure \ref{single_beta} which plots the induced priors on $v$ based on common choices of priors on $a_0$, and Figure \ref{single_IG} which plots the induced priors on $a_0$ based on common choices of priors on $v$. The following results provide a quick reference for common induced priors we will get using Theorem \ref{single}. 

Applying the formulas provided in Theorem \ref{single}, we plot the induced priors on $v$ based on example choices of prior on $a_0$ (e.g., the uniform prior) in the NPP in Figure \ref{single_beta}. The distributions beta(1,1), beta(0.5,0.5), beta(2,10) and beta(10,2) distributions are considered for the prior for $a_0$ in the NPP. The figure presents the inverse gamma prior that matches most closely to the prior for $v$ induced by the prior on $a_0$ based on the formula in Theorem \ref{single}. Specifically, we solve for the inverse gamma distribution which  minimizes the Kullback–Leibler (KL) divergence from the induced prior on $v$. The resulting induced priors on $v$ are IG($0.3$,$10^{-3}$), IG($0.1$,$10^{-5}$), IG($1.7$,$0.2$) and IG($1.2$,$0.003$).
We observe that as the prior on $a_0$ encourages more borrowing of historical information, i.e. as the expectation of the beta distribution increases, the expectation of the prior on $v$ decreases, i.e., the prior on $v$ correspondingly encourages more borrowing.
For example, as illustrated by the two plots in the second row, in the left plot, a beta(2,10) prior on $a_0$ is concentrated near zero, indicating that borrowing is discouraged, and the corresponding prior on $v$ has relatively large variance.
In the right plot, a beta(10,2) prior on $a_0$ is concentrated near one, indicating that borrowing is encouraged, and the corresponding prior on $v$ is much more concentrated near zero, also indicating that borrowing is encouraged. In Figure \ref{IG} in Appendix \ref{appen:fig}, we provide similar plots for three beta priors for $a_0$ each having mean equal to $0.5$ in the NPP, and derive the corresponding inverse gamma distribution prior for $v$ in a BHM.

Conversely, we plot the priors induced on $a_0$ based on example choices of prior on $v$ from the inverse gamma family in the BHM in Figure \ref{single_IG}. We choose IG(3,10), IG(3,1), IG(1,0.1), IG(1,0.01) as the prior for $v$ in the BHM, and find the beta prior that most closely matches the induced prior on $a_0$ based on the formula in Theorem \ref{single}. 
Specifically, we draw samples of $v$ from the inverse gamma distributions and transform them to samples of $a_0$.
We then use beta regression to solve for the closest beta distribution that fits the $a_0$ samples. The resulting induced priors on $a_0$ are beta($3$,$404$), beta($3$,$44$), beta($1.2$,$5.5$) and beta($1.9$,$1.5$). 
In the first row, in the left plot, the IG(3,10) distribution has mean equal to 5 and variance equal to 25, and the corresponding beta distribution is highly  concentrated near zero, indicating that borrowing is discouraged.
By comparison, in the right plot, the IG(3,1) distribution has mean equal to 0.5 and variance equal to 0.25, supporting comparatively more borrowing, and the corresponding prior on $a_0$ is much less concentrated near zero.

In summary, we have provided a reference for common induced priors we will obtain using Theorem \ref{single}, as well as some visual intuition on the connection between the prior on $a_0$ for the NPP and the prior on $v$ for the BHM. The significance of establishing this connection is two-fold. It sheds light on the choice of $a_0$ in the NPP from the perspective of the canonical BHM. Moreover, when deciding on how much information to borrow, calibrating a parameter on the scale of $0 < a_0 < 1$ is more intuitive than calibrating a variance parameter. For example, a beta(10,2) prior on $a_0$ is easier to grasp than an IG(1.2, 0.003) prior on $v$ which has undefined variance. Since these two prior choices are shown to lead to equivalent posterior inference on $\theta$, in the context of adjusting information borrowing, using the NPP has the advantage of easier prior calibration compared to the BHM. 

\section{Analysis of Pediatric Lupus Trial}

We now demonstrate the use of the ANPP through an important application in a pediatric lupus clinical trial, where historical data from an adult phase 3 program is publicly available.
Borrowing information from adult trials, so-called, partial extrapolation, is an increasingly common strategy in pediatric drug development due to the challenges in enrolling pediatric patients in clinical trials and the ability of regulators to require trials in these populations \citep{fda_guide}.
The enrollment of patients in pediatric trials is often challenging due to the limited number of available patients, parental safety concerns, and technical limitations \citep{greenberg_2018}.
Utilizing Bayesian methods for extrapolating adult data in pediatric trials through informative priors is a natural approach, as illustrated in the FDA guidance on complex innovative designs \citep{fda}.

Belimumab (Benlysta) is a biologic for the treatment of adults with active, systemic lupus erythematosus (SLE).
As a part of the pediatric investigation plan, the PLUTO clinical trial \citep{ped_2020} was conducted to evaluate the effect of belimumab on children aged 5 to 17 with active, seropositive SLE who receive standard therapy.
Previous phase 3 trials, BLISS-52 and BLISS-76 \citep{Furie_2011, Navarra_2011}, established the efficacy of belimumab alongside standard therapy for adults. The FDA review of the PLUTO trial submission utilized data from these adult trials to inform the approval decision.
All three trials shared the same composite primary outcome, the SLE Responder Index (SRI-4).

\begin{figure}[H]
\begin{center}
\includegraphics[width=0.9\textwidth]{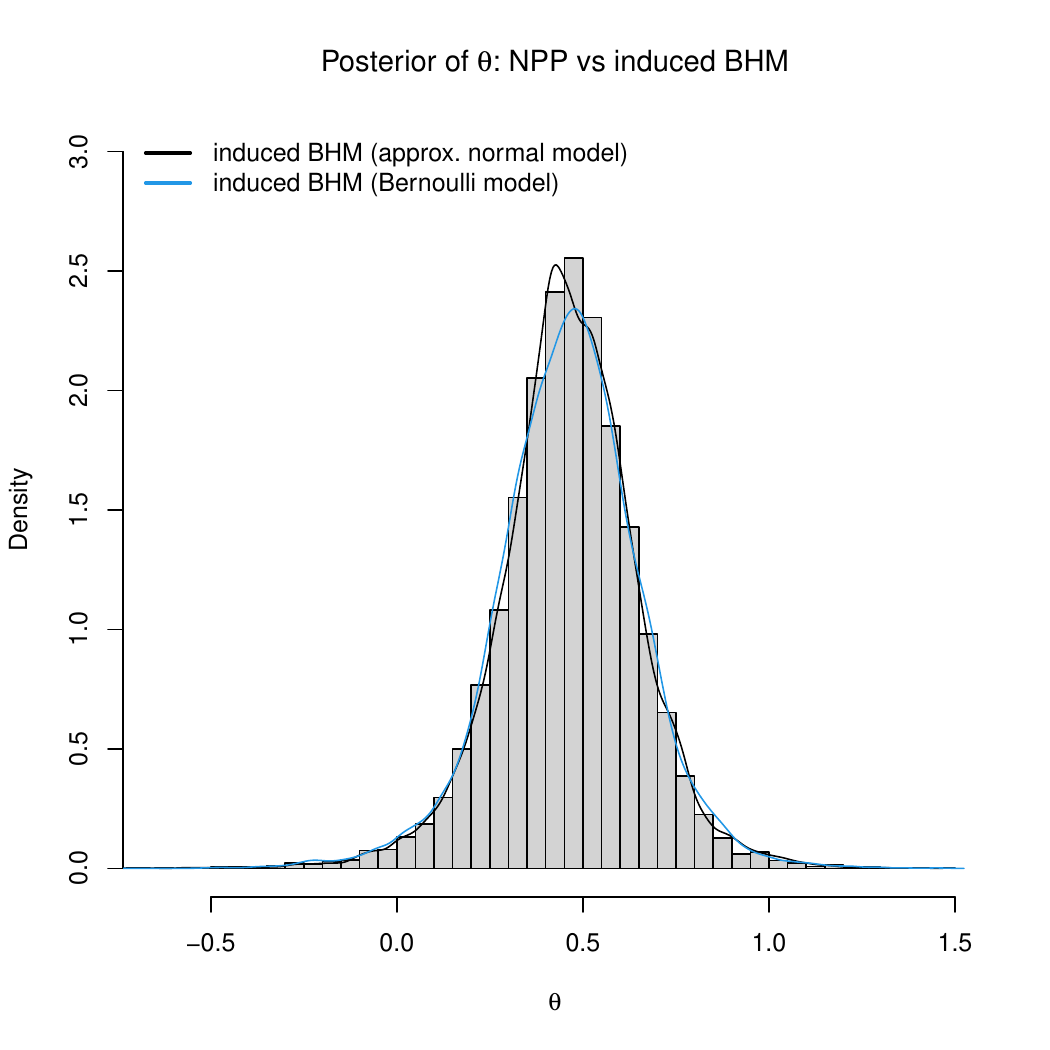}
\captionsetup{width=0.8\linewidth}
\caption{Pediatric lupus trial: the histogram represents the posterior of $\theta$ using a ANPP with a uniform prior on $a_0$ and $h_k(a_0)$ chosen according to Theorem \ref{multiple}. The black density curve represents the posterior of $\theta$ using the BHM where the prior on $v$ is induced using Theorem \ref{multiple}. We observe that the two posteriors are equivalent. They are also equivalent to the posterior of $\theta$ using a Bernoulli model (blue curve) for the induced BHM. We ran four independent chains of 10,000 iterations with 5,000 burn-ins using RStan.}
\label{ped_hist}
\end{center}
\end{figure}

We perform a Bayesian analysis of the PLUTO study data ($n=92$), incorporating information from adult studies BLISS-52 ($n_{01}=548$) and BLISS-76 ($n_{02}=577$)  using the ANPP and the iNPP.
Since Theorem \ref{multiple} requires i.i.d. normal data but our trial outcomes are binary, the theorem does not directly apply without approximation.
Here, we assume the log odds ratio of receiving treatment is approximately normally distributed.
Let $n_t$ and $n_c$ denote the number of patients in the treatment and control group in the PLUTO study, respectively. Let $p_t$ and $p_c$ denote the probability of response to treatment in the treatment and control group, respectively.
The parameter of interest is the log odds ratio, denoted by $\theta$. It can be estimated by $$\hat{\theta} = \log\left(\frac{\hat{p}_t}{1-\hat{p}_t}\right) - \log\left(\frac{\hat{p}_c}{1-\hat{p}_c}\right),$$ with asymptotic variance given by $$Var(\hat{\theta}) = \frac{1}{n_tp_t} + \frac{1}{n_t(1-p_t)} + \frac{1}{n_cp_c} + \frac{1}{n_c(1-p_c)}.$$ The approximate mean and variance of $\theta$ for the two adult studies can be computed analogously.
For each current and historical dataset, we construct an approximation normal likelihood with sample size one, mean $\hat{\theta}$ and known variance Var($\hat{\theta}$), so that Theorem \ref{multiple} can be applied. Let $a_{01}$ and $a_{02}$ denote the discounting parameters for BLISS-52 and BLISS-76, respectively.
We assume independent uniform priors for $a_{01}$ and $a_{02}$ for the iNPP. 
For the ANPP, $a_{01}=h_1(a_0)$ and $a_{02}=h_2(a_0)$, and we use the induced prior on $v$ by Theorem \ref{multiple} where $\pi(a_0)$ is the uniform distribution. We provide Figure \ref{ped_hist} below that corroborates empirically that the resulting posteriors for $\theta$ based on the BHM and the ANPP using the approximate normal model are equivalent, and they are also equivalent to the posterior for $\theta$ using a Bernoulli model for the induced BHM.

\begin{figure}[H]
\begin{center}
\includegraphics[width=0.9\textwidth]{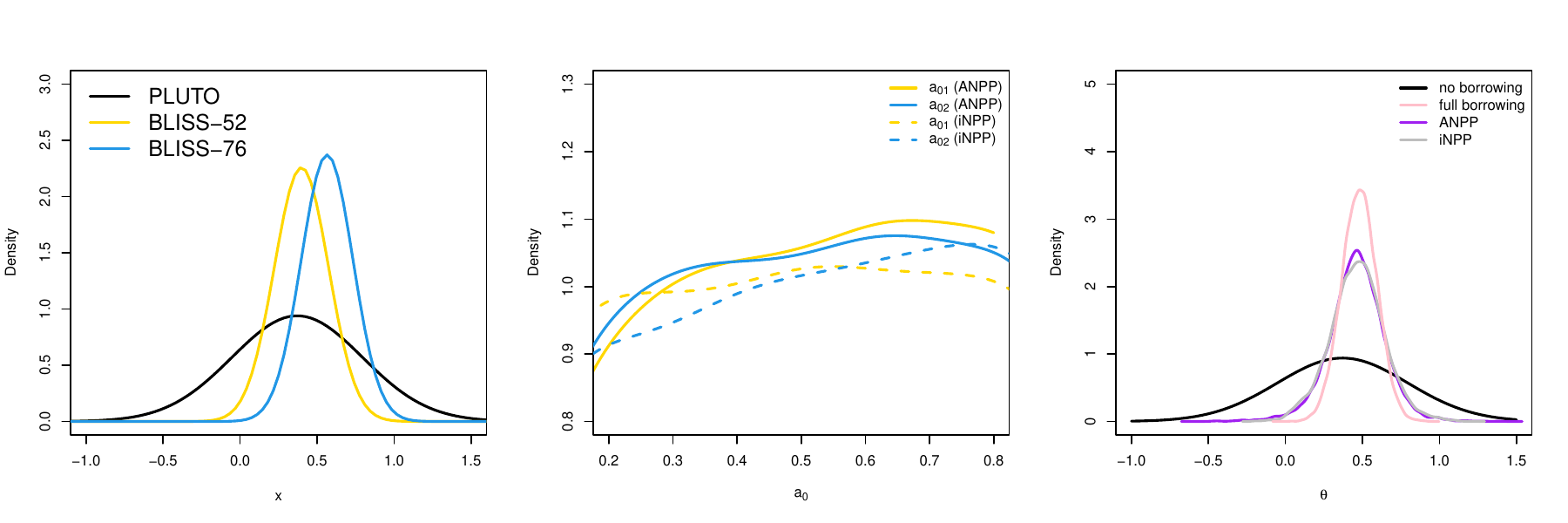}
\captionsetup{width=0.8\linewidth}
\caption{Marginal posteriors for $a_{01}$, $a_{02}$ and $\theta$ using the ANPP and the iNPP for the pediatric lupus study.
The first column includes plots of the approximate normal densities of the log odds ratio of the PLUTO study (black line) and the two adult studies (yellow and blue lines).
The second column includes plots of the posterior densities of $a_{01}$ and $a_{02}$ using the ANPP (yellow and blue solid lines) and the iNPP (yellow and blue dashed lines).
The third column includes plots of the posterior densities of $\theta$ using four different priors, the ANPP (purple), the iNPP (grey), the power prior with $a_0=0$ (black) and the power prior with $a_0=1$ (pink). The ANPP and iNPP are implemented with RStan, and the full borrowing case (power prior with $a_0=1$) is implemented with BayesPPD.}
\label{ped_post}
\end{center}
\end{figure}

Figure \ref{ped_post} shows the approximate normal likelihoods of the log odds ratio for the three datasets and the marginal posteriors for $a_{01}$, $a_{02}$ and $\theta$ for the ANPP and iNPP. The ANPP and iNPP are implemented with RStan, and the full borrowing case (power prior with $a_0=1$) is implemented with BayesPPD.
We can see that the estimates of the log odds ratios are quite similar for the adult studies and the pediatric study, although the variance is much smaller for the adult studies due to their larger sample sizes.
The posteriors for $a_{01}$ and $a_{02}$ using ANPP and iNPP are similar, and they both have modes at one.
The ANPP leads to more borrowing from the smaller dataset, while the iNPP leads to more borrowing from the larger dataset.
The posteriors for $\theta$ using the two models are almost identical. 
The posterior means using the ANPP and iNPP are similar to that of the no borrowing model, but the posterior variances are much smaller.
The posterior mean, standard deviation, and 95\% credible interval for $\theta$, $a_{01}$ and $a_{02}$ for the ANPP and iNPP are displayed in Table \ref{ped_theta}.
The posterior means of $\theta$ are similar for the two models.
The posterior means of $a_{01}$ and $a_{02}$ using the ANPP are slightly higher than those of the iNPP, indicating more information borrowed.

\begin{table}
\caption{Pediatric lupus trial: posterior mean, standard deviation, and 95\% credible interval for $\theta$, $a_{01}$ and $a_{02}$}
\begin{center}
\begin{tabular}{lccc}
\hline
Parameter & Mean & SD & 95\% Credible Interval  \\[5pt]
\hline
Adapted NPP&&&\\
\hline
$\theta$ & 0.46 & 0.18 & (0.08, 0.81)    \\
$a_{01}$ & 0.53 & 0.28 & (0.05, 0.98)    \\
$a_{02}$ & 0.52 & 0.28 & (0.05, 0.98)    \\
\hline
Independent NPP&&&\\
\hline
$\theta$ & 0.47 & 0.18 & (0.10, 0.82)    \\
$a_{01}$ & 0.51 & 0.29 & (0.26, 0.97)   \\
$a_{02}$ & 0.52 & 0.29 & (0.28, 0.97)    \\
\hline
\end{tabular}
\end{center}
\label{ped_theta}
\end{table}

\section{Discussion} 

In this paper, we have developed the adapted normalized power prior (ANPP) for multiple historical datasets. We have established a direct relationship between the priors for the discounting parameters of the ANPP and the prior for the variance parameter of the BHM that leads to equivalent posterior inference for the normal model.
Establishing this relationship gives us semi-automatic benchmark priors on the discounting parameters from the perspective of the BHM. The ANPP allows for dependent discounting of the historical datasets without the need to specify a prior for each discounting parameter, since each discounting parameter is shown to be a deterministic function of a single global discounting parameter. In addition, in the context of adjusting the amount of information borrowed, the calibration of a prior on a parameter on the scale of $0 < a_0 < 1$ is more intuitive than that of a variance parameter. Using the ANPP has the advantage of easier prior calibration compared to the BHM.

Based on our simulations, we observed that the ANPP and iNPP behave similarly when the historical and current datasets are compatible.
The ANPP is more sensitive to conflicts in the data than the iNPP, and it tends to borrow more from the smaller dataset.
If the historical datasets vary in the degree of compatibility, the iNPP borrows more from the compatible historical datasets while the ANPP discounts based on the overall heterogeneity across all datasets. In the pediatric trial analysis, we have shown how Theorem \ref{multiple} can be applied more generally for non-normal data using asymptotic approximations.

Based on Theorems \ref{single} and \ref{multiple}, we can see that the beta prior on $a_0$ in an NPP does not generally translate perfectly to an inverse gamma prior on $v$ in a BHM, and thus for an NPP or ANPP to be truly equivalent to a BHM, one will need to consider non-standard priors as a general rule. Future research will investigate how Theorems \ref{single} and \ref{multiple} can be extended to non-normal i.i.d. data as well as generalized linear models.

\bibliographystyle{chicago}
\bibliography{refs}

\begin{thebibliography}{}

\bibitem[\protect\citeauthoryear{Banbeta, van Rosmalen, Dejardin, and
  Lesaffre}{Banbeta et~al.}{2019}]{Banbeta_2019}
Banbeta, A., J.~van Rosmalen, D.~Dejardin, and E.~Lesaffre (2019, Mar).
\newblock Modified power prior with multiple historical trials for binary
  endpoints.
\newblock {\em Statistics in Medicine\/}~{\em 38\/}(7), 1147–1169.

\bibitem[\protect\citeauthoryear{Bennett, White, Best, and Mander}{Bennett
  et~al.}{2021}]{Bennett_2021}
Bennett, M., S.~White, N.~Best, and A.~Mander (2021).
\newblock A novel equivalence probability weighted power prior for using
  historical control data in an adaptive clinical trial design: A comparison to
  standard methods.
\newblock {\em Pharmaceutical Statistics\/}~{\em 20\/}(3), 462--484.

\bibitem[\protect\citeauthoryear{Berry, Broglio, Groshen, and Berry}{Berry
  et~al.}{2013}]{berry}
Berry, S., K.~Broglio, S.~Groshen, and D.~Berry (2013).
\newblock {B}ayesian hierarchical modeling of patient subpopulations: efficient
  designs of phase ii oncology clinical trials.
\newblock {\em Clin Trials\/}~{\em 10}, 720--734.

\bibitem[\protect\citeauthoryear{Brunner, Abud-Mendoza, Viola, Calvo~Penades,
  Levy, Anton, Calderon, Chasnyk, Ferrandiz, Keltsev, and et~al.}{Brunner
  et~al.}{2020}]{ped_2020}
Brunner, H.~I., C.~Abud-Mendoza, D.~O. Viola, I.~Calvo~Penades, D.~Levy,
  J.~Anton, J.~E. Calderon, V.~G. Chasnyk, M.~A. Ferrandiz, V.~Keltsev, and
  et~al. (2020, Jul).
\newblock Safety and efficacy of intravenous belimumab in children with
  systemic lupus erythematosus: results from a randomised, placebo-controlled
  trial.
\newblock {\em Annals of the Rheumatic Diseases\/}~{\em 79}, 1340--1348.

\bibitem[\protect\citeauthoryear{Carvalho and Ibrahim}{Carvalho and
  Ibrahim}{2021}]{Carvalho_Ibrahim_2021}
Carvalho, L.~M. and J.~G. Ibrahim (2021, Jul).
\newblock On the normalized power prior.
\newblock {\em Statistics in Medicine\/}~{\em 40\/}(24), 5251--5275.

\bibitem[\protect\citeauthoryear{Chen and Ibrahim}{Chen and
  Ibrahim}{2006}]{chen_2006}
Chen, M.-H. and J.~Ibrahim (2006, 09).
\newblock The relationship between the power prior and hierarchical models.
\newblock {\em Bayesian Analysis\/}~{\em 1}.

\bibitem[\protect\citeauthoryear{Duan, Ye, and Smith}{Duan
  et~al.}{2006}]{duan_2006}
Duan, Y., K.~Ye, and E.~P. Smith (2006, feb).
\newblock Evaluating water quality using power priors to incorporate historical
  information.
\newblock {\em Environmetrics (London, Ont.)\/}~{\em 17\/}(1), 95--106.

\bibitem[\protect\citeauthoryear{Furie, Petri, Zamani, Cervera, Wallace,
  Tegzová, Sanchez-Guerrero, Schwarting, Merrill, Chatham, and et~al.}{Furie
  et~al.}{2011}]{Furie_2011}
Furie, R., M.~Petri, O.~Zamani, R.~Cervera, D.~J. Wallace, D.~Tegzová,
  J.~Sanchez-Guerrero, A.~Schwarting, J.~T. Merrill, W.~W. Chatham, and et~al.
  (2011, Dec).
\newblock A phase {III}, randomized, placebo-controlled study of belimumab, a
  monoclonal antibody that inhibits {B} lymphocyte stimulator, in patients with
  systemic lupus erythematosus.
\newblock {\em Arthritis and Rheumatism\/}~{\em 63\/}(12), 3918–3930.

\bibitem[\protect\citeauthoryear{Gelman, Carlin, Stern, Dunson, Vehtari, and
  Rubin}{Gelman et~al.}{2013}]{gelman}
Gelman, A., J.~Carlin, H.~Stern, D.~Dunson, A.~Vehtari, and D.~Rubin (2013).
\newblock {\em Bayesian Data Analysis, Third Edition}.
\newblock Chapman \& Hall/CRC Texts in Statistical Science. Taylor \& Francis.

\bibitem[\protect\citeauthoryear{Gravestock and Held}{Gravestock and
  Held}{2019}]{Gravestock_Held_2019}
Gravestock, I. and L.~Held (2019).
\newblock Power priors based on multiple historical studies for binary
  outcomes.
\newblock {\em Biometrical Journal\/}~{\em 61\/}(5), 1201–1218.

\bibitem[\protect\citeauthoryear{Gravestock, Held, and consortium}{Gravestock
  et~al.}{2017}]{Gravestock_Held_2017}
Gravestock, I., L.~Held, and C.-N. consortium (2017, Jun).
\newblock Adaptive power priors with empirical {B}ayes for clinical trials.
\newblock {\em Pharmaceutical Statistics\/}~{\em 16\/}(5), 349--360.

\bibitem[\protect\citeauthoryear{Greenberg, Gamel, Bloom, Bradley, Jafri,
  Hinton, Nambiar, Wheeler, Tiernan, Smith, and et~al.}{Greenberg
  et~al.}{2018}]{greenberg_2018}
Greenberg, R.~G., B.~Gamel, D.~Bloom, J.~Bradley, H.~S. Jafri, D.~Hinton,
  S.~Nambiar, C.~Wheeler, R.~Tiernan, P.~B. Smith, and et~al. (2018, Mar).
\newblock Parents’ perceived obstacles to pediatric clinical trial
  participation: Findings from the clinical trials transformation initiative.
\newblock {\em Contemporary clinical trials communications\/}~{\em 9}, 33–39.

\bibitem[\protect\citeauthoryear{Han, Ye, and Wang}{Han
  et~al.}{2022}]{Han_Ye_Wang_2022}
Han, Z., K.~Ye, and M.~Wang (2022, Mar).
\newblock A study on the power parameter in power prior {B}ayesian analysis.
\newblock {\em The American Statistician\/}, 1--8.

\bibitem[\protect\citeauthoryear{Han, Zhang, Tiwari, and Bai}{Han
  et~al.}{2025}]{han_2025}
Han, Z., Q.~Zhang, R.~Tiwari, and T.~Bai (2025).
\newblock Bayesian borrowing with multiple heterogeneous historical studies
  using order restricted normalized power prior.
\newblock {\em Statistics in Medicine\/}~{\em 44\/}(3-4), e10302.

\bibitem[\protect\citeauthoryear{Ibrahim and Chen}{Ibrahim and
  Chen}{2000}]{chen_2000}
Ibrahim, J.~G. and M.-H. Chen (2000, feb).
\newblock Power prior distributions for regression models.
\newblock {\em Statistical Science\/}~{\em 15\/}(1), 46--60.

\bibitem[\protect\citeauthoryear{Liu}{Liu}{2018}]{Liu_2018}
Liu, G.~F. (2018).
\newblock A dynamic power prior for borrowing historical data in noninferiority
  trials with binary endpoint.
\newblock {\em Pharmaceutical Statistics\/}~{\em 17\/}(1), 61–73.

\bibitem[\protect\citeauthoryear{Navarra, Guzmán, Gallacher, Hall, Levy,
  Jimenez, Li, Thomas, Kim, León, and et~al.}{Navarra
  et~al.}{2011}]{Navarra_2011}
Navarra, S.~V., R.~M. Guzmán, A.~E. Gallacher, S.~Hall, R.~A. Levy, R.~E.
  Jimenez, E.~K.-M. Li, M.~Thomas, H.-Y. Kim, M.~G. León, and et~al. (2011,
  Feb).
\newblock Efficacy and safety of belimumab in patients with active systemic
  lupus erythematosus: a randomised, placebo-controlled, phase 3 trial.
\newblock {\em The Lancet\/}~{\em 377\/}(9767), 721–731.

\bibitem[\protect\citeauthoryear{Neuenschwander, Branson, and
  Spiegelhalter}{Neuenschwander et~al.}{2009}]{neuens_2009}
Neuenschwander, B., M.~Branson, and D.~J. Spiegelhalter (2009).
\newblock A note on the power prior.
\newblock {\em Statistics in Medicine\/}~{\em 28}, 3562--3566.

\bibitem[\protect\citeauthoryear{Pan, Yuan, and Xia}{Pan
  et~al.}{2017}]{Pan_Yuan_Xia_2017}
Pan, H., Y.~Yuan, and J.~Xia (2017, Nov).
\newblock A calibrated power prior approach to borrow information from
  historical data with application to biosimilar clinical trials.
\newblock {\em Journal of the Royal Statistical Society. Series C, Applied
  statistics\/}~{\em 66\/}(5), 979–996.

\bibitem[\protect\citeauthoryear{Pawel, Aust, Held, and Wagenmakers}{Pawel
  et~al.}{2023a}]{pawel_2023}
Pawel, S., F.~Aust, L.~Held, and E.-J. Wagenmakers (2023a).
\newblock Normalized power priors always discount historical data.
\newblock {\em Stat\/}~{\em 12\/}(1), e591.

\bibitem[\protect\citeauthoryear{Pawel, Aust, Held, and Wagenmakers}{Pawel
  et~al.}{2023b}]{pawel_test}
Pawel, S., F.~Aust, L.~Held, and E.-J. Wagenmakers (2023b).
\newblock Power priors for replication studies.
\newblock {\em TEST\/}.

\bibitem[\protect\citeauthoryear{Psioda and Xue}{Psioda and
  Xue}{2020}]{psioda_ped}
Psioda, M.~A. and X.~Xue (2020).
\newblock A {B}ayesian adaptive two-stage design for pediatric clinical trials.
\newblock {\em Journal of Biopharmaceutical Statistics\/}~{\em 30\/}(6),
  1091--1108.

\bibitem[\protect\citeauthoryear{Shen, Carvalho, Psioda, and Ibrahim}{Shen
  et~al.}{2024}]{shen_2023}
Shen, Y., L.~M. Carvalho, M.~A. Psioda, and J.~G. Ibrahim (2024).
\newblock Optimal priors for the discounting parameter of the normalized power
  prior.
\newblock {\em Statistica Sinica\/}.
\newblock (in press).

\bibitem[\protect\citeauthoryear{Shen, Psioda, and Ibrahim}{Shen
  et~al.}{2023a}]{shen_RJ}
Shen, Y., M.~A. Psioda, and J.~G. Ibrahim (2023a).
\newblock {B}ayes{PPD}: An {R} package for {B}ayesian sample size determination
  using the power and normalized power prior for generalized linear models.
\newblock {\em The R Journal\/}~{\em 14}, 335--351.
\newblock https://doi.org/10.32614/RJ-2023-016.

\bibitem[\protect\citeauthoryear{Shen, Psioda, and Ibrahim}{Shen
  et~al.}{2023b}]{bayesppd}
Shen, Y., M.~A. Psioda, and J.~G. Ibrahim (2023b).
\newblock {\em BayesPPD: Bayesian Power Prior Design}.
\newblock R package version 1.1.2.

\bibitem[\protect\citeauthoryear{{Stan Development Team}}{{Stan Development
  Team}}{2024}]{rstan}
{Stan Development Team} (2024).
\newblock {RStan}: the {R} interface to {Stan}.
\newblock R package version 2.32.6.

\bibitem[\protect\citeauthoryear{{U.S. Food and Drug Administration}}{{U.S.
  Food and Drug Administration}}{2016}]{fda_guide}
{U.S. Food and Drug Administration} (2016).
\newblock {\em Leveraging existing clinical data for extrapolation to pediatric
  uses of medical devices: guidance for industry and Food and Drug
  Administration Staff}.

\bibitem[\protect\citeauthoryear{{U.S. Food and Drug Administration}}{{U.S.
  Food and Drug Administration}}{2019}]{fda}
{U.S. Food and Drug Administration} (2019).
\newblock {\em Interacting with the {FDA} on complex innovative trial designs
  for drugs and biological products: draft guidance for industry}.

\bibitem[\protect\citeauthoryear{{U.S. Food and Drug Administration}}{{U.S.
  Food and Drug Administration}}{2022}]{fda_e11a}
{U.S. Food and Drug Administration} (2022).
\newblock {\em {ICH} Harmonised Guideline Pediatric Extrapolation {E11A}}.

\bibitem[\protect\citeauthoryear{Ye, Han, Duan, and Bai}{Ye
  et~al.}{2022}]{YE202229}
Ye, K., Z.~Han, Y.~Duan, and T.~Bai (2022).
\newblock Normalized power prior {B}ayesian analysis.
\newblock {\em Journal of Statistical Planning and Inference\/}~{\em 216},
  29--50.

\end{thebibliography}

\appendix

\section{Proofs from Section 2}

\subsection{Proof of Theorem \ref{single}}\label{appen:th1}
\begin{proof}\
We first simplify the joint posterior of $\theta$ and $a_0$ under the NPP
\begin{align*}
\pi(\theta,a_0|y_0) &\propto \exp \left\{-\frac{1}{2 \sigma^{2}} \sum_{i=1}^{n}\left(y_{i}-\theta\right)^{2}\right\}\frac{\left[\sigma_0^{-n_{0}} \exp \left\{-\frac{1}{2 \sigma_{0}^{2}} \sum_{i=1}^{n_{0}}\left(y_{0i}-\theta\right)^{2}\right\}\right]^{a_{0}} }{\int \left[\sigma_{0}^{-n_{0}} \exp \left\{-\frac{1}{2 \sigma_{0}^{2}} \sum_{i=1}^{n_{0}}\left(y_{0i}-\theta\right)^{2}\right\}\right]^{a_{0}} d\theta}\pi(a_0),\\
&\propto \exp \left\{-\frac{1}{2 \sigma^{2}} \sum_{i=1}^{n}\left(y_{i}-\theta\right)^{2}\right\}\frac{\left[\sigma_0^{-n_{0}} \exp \left\{-\frac{1}{2 \sigma_{0}^{2}} \sum_{i=1}^{n_{0}}\left(y_{0i}-\theta\right)^{2}\right\}\right]^{a_{0}} }{\left(\frac{2\pi\sigma_0^2}{n_0a_0}\right)^{\frac{1}{2}} \exp\{-\frac{n_0a_0}{2\sigma_0^2}(-\bar{y_{0}}^2 + \sum_{i=1}^{n_0}y_{0i}^2/n_0)\}}\pi(a_0).\\
\end{align*}
The power prior part simplifies to 
\begin{align*}
&\exp \left\{-\frac{1}{2 \sigma^{2}} \sum_{i=1}^{n}\left(y_{i}-\theta\right)^{2}\right\}\left[\sigma_0^{-n_{0}} \exp \left\{-\frac{1}{2 \sigma_{0}^{2}} \sum_{i=1}^{n_{0}}\left(y_{0i}-\theta\right)^{2}\right\}\right]^{a_{0}},\\
\propto & N(\mu_p, \sigma_p^2)\sigma_p\exp\left\{\frac{1}{2}\left(\frac{n}{\sigma^2}+\frac{a_0n_0}{\sigma_0^2}\right)\left(\frac{\frac{n\bar{y}}{\sigma^2}+\frac{a_0n_0\bar{y}_0}{\sigma_0^2}}{\frac{n}{\sigma^2}+\frac{a_0n_0}{\sigma_0^2}}\right)^2\right\}\exp\left\{-\frac{a_0}{2\sigma_0^2}\sum_{i=1}^{n_0}y_{0i}^2\right\},
\end{align*}
where $N(m, v)$ is the probability density function of a normal distribution with parameters $m$ and $v$ and $\mu_p=\frac{\frac{n\bar{y}}{\sigma^2}+\frac{a_0n_0\bar{y}_0}{\sigma_0^2}}{\frac{n}{\sigma^2}+\frac{a_0n_0}{\sigma_0^2}}$, $\sigma_p^2=\left(\frac{n}{\sigma^2}+\frac{a_0n_0}{\sigma_0^2}\right)^{-1}$.
Then the joint posterior of $\theta$ and $a_0$ under the NPP is 
\begin{align*}
\pi(\theta, a_0|y, y_0) &\propto \exp\left\{\frac{\mu_p^2}{2\sigma_p^2}\right\}\sigma_p\exp\left\{-\frac{a_0}{2\sigma_0^2}\sum_{i=1}^{n_0}y_{0i}^2 -\frac{n_0a_0}{2\sigma_0^2}\bar{y_{0}}^2 + \frac{a_0}{2\sigma_0^2}\sum_{i=1}^{n_0}y_{0i}^2\right\}a_0^{\frac{1}{2}}N(\mu_p, \sigma_p^2)\pi(a_0),\\
&\propto \sigma_p\exp\left\{\frac{\mu_p^2}{2\sigma_p^2} -\frac{n_0a_0}{2\sigma_0^2}\bar{y_{0}}^2\right\}a_0^{\frac{1}{2}}N(\mu_p, \sigma_p^2)\pi(a_0).\\
\end{align*}
We now proceed to derive the joint posterior of $\theta$ and $v$ with a BHM. The joint posterior of $\theta$, $\theta_0$, $\mu$, and $v$ with the BHM is 
\begin{tiny}
\begin{align*}
\pi(\theta,\theta_0,\mu,v|y, y_0) \propto \pi(v)\frac{1}{v}\exp\left\{-\frac{n(\bar{y}-\theta)^2}{2\sigma^2} - \frac{n_0(\bar{y}_0-\theta_0)^2}{2\sigma_0^2}-\frac{(\theta-\mu)^2+(\theta_0-\mu)^2}{2v}-\frac{(\mu-\alpha)^2}{2\nu^2}\right\}\pi(v).
\end{align*}
\end{tiny}
Integrating out $\mu$, we get 
\begin{align*}
\pi(\theta,\theta_0,v|y, y_0) \propto & \hspace{2mm}\pi(v)\frac{1}{v} \exp\left\{-\frac{n(\bar{y}-\theta)^2}{2\sigma^2} - \frac{n_0(\bar{y}_0-\theta_0)^2}{2\sigma_0^2}\right\} \times \\
&\int\exp\left\{-\frac{(\theta-\mu)^2+(\theta_0-\mu)^2}{2v}-\frac{(\mu-\alpha)^2}{2\nu^2}\right\}d\mu.
\end{align*}
Since
\begin{align*}
&\int\exp\left\{-\frac{(\theta-\mu)^2+(\theta_0-\mu)^2}{2v}-\frac{(\mu-\alpha)^2}{2\nu^2}\right\}d\mu,\\
&\propto \exp\left\{-\frac{1}{2v}(\theta^2+\theta_0^2)\right\}\left(\frac{2}{v}+\frac{1}{\nu^2}\right)^{-\frac{1}{2}}\exp\left\{\frac{1}{2}\left(\frac{2}{v}+\frac{1}{\nu^2}\right)\left(\frac{\frac{\theta+\theta_0}{v}+\frac{\alpha}{\nu^2}}{\frac{2}{v}+\frac{1}{\nu^2}}\right)^2\right\},
\end{align*}
Then 
\begin{tiny}
\begin{align*}
\pi(\theta,\theta_0,v|y, y_0) &\propto \pi(v)\frac{1}{v}\left(\frac{2}{v}+\frac{1}{\nu^2}\right)^{-\frac{1}{2}} \times\\
& \exp\left\{-\frac{n(\bar{y}-\theta)^2}{2\sigma^2} - \frac{n_0(\bar{y}_0-\theta_0)^2}{2\sigma_0^2}-\frac{1}{2v}(\theta^2+\theta_0^2)+\frac{1}{2}\left(\frac{2}{v}+\frac{1}{\nu^2}\right)\left(\frac{\frac{\theta+\theta_0}{v}+\frac{\alpha}{\nu^2}}{\frac{2}{v}+\frac{1}{\nu^2}}\right)^2\right\}.
\end{align*}
\end{tiny}
Next, integrating out $\theta_0$ gives 
\begin{align*}
\pi(\theta,\theta_0,v|y, y_0) &\propto \pi(v)\frac{1}{v}\left(\frac{2}{v}+\frac{1}{\nu^2}\right)^{-\frac{1}{2}}\exp\left\{-\frac{n(\bar{y}-\theta)^2}{2\sigma^2}-\frac{\theta^2}{2v}\right\}\times\\
& \int\exp\left\{-\frac{n_0(\bar{y}_0-\theta_0)^2}{2\sigma_0^2}-\frac{\theta_0^2}{2v}+\frac{1}{2}\left(\frac{2}{v}+\frac{1}{\nu^2}\right)\left(\frac{\frac{\theta+\theta_0}{v}+\frac{\alpha}{\nu^2}}{\frac{2}{v}+\frac{1}{\nu^2}}\right)^2\right\}d\theta_0.
\end{align*}
The integral term is proportional to 
\begin{align*}
&\exp\left\{\frac{1}{2}\left[\frac{(\theta/v)^2+2\theta\alpha/(v\nu^2)+\alpha^2/(\nu^4)}{\frac{2}{v}+\frac{1}{\nu^2}}\right]\right\} \times\\
&\int\exp\left\{\frac{n_0\bar{y}_0\theta_0}{\sigma_0^2}-\frac{n_0\theta_0^2}{2\sigma_0^2}-\frac{\theta_0^2}{2v} + \frac{\frac{1}{2}[2\theta\theta_0/(v^2) + (\theta_0/v)^2 + 2\theta_0\alpha/(v\nu^2)]}{\frac{2}{v}+\frac{1}{\nu^2}}\right\}d\theta_0\\
=&\exp\left\{\frac{1}{2}\left[\frac{(\theta/v)^2+2\theta\alpha/(v\nu^2)+\alpha^2/(\nu^4)}{\frac{2}{v}+\frac{1}{\nu^2}}\right]\right\} \times\\
&\left(\frac{n_0}{\sigma_0^2}+\frac{1}{v}-\frac{1/(v^2)}{\frac{2}{v}+\frac{1}{\nu^2}}\right)^{-\frac{1}{2}}\exp\left\{\frac{1}{2}\frac{\left(\frac{n_0\bar{y}_0}{\sigma_0^2} + \frac{\theta/(v^2)}{\frac{2}{v}+\frac{1}{\nu^2}} + \frac{\alpha/(v\nu^2)}{\frac{2}{v}+\frac{1}{\nu^2}}\right)^2}{\frac{n_0}{\sigma_0^2}+\frac{1}{v}-\frac{1/(v^2)}{\frac{2}{v}+\frac{1}{\nu^2}}}\right\}.
\end{align*}
For simplicity of notation, denote $U = \frac{2}{v}+\frac{1}{\nu^2}$ and $W = \frac{n_0}{\sigma_0^2}+\frac{1}{v}-\frac{1/(v^2)}{\frac{2}{v}+\frac{1}{\nu^2}}$.\\
Collecting all the terms involving $\theta$ and $v$, we get
\begin{tiny}
\begin{align*}
\pi(\theta, v|y, y_0) \propto &\hspace{2mm}\pi(v)\frac{1}{v}U^{-\frac{1}{2}}\exp\left\{-\frac{n(\bar{y}-\theta)^2}{2\sigma^2}-\frac{\theta^2}{2v}\right\}\exp\left\{\frac{1}{2}\left[\frac{(\theta/v)^2+2\theta\alpha/(v\nu^2)+\alpha^2/(\nu^4)}{U}\right]\right\} \times \\
&W^{-\frac{1}{2}}\exp\left\{\frac{1}{2}\frac{\left(\frac{n_0\bar{y}_0}{\sigma_0^2} + \frac{\theta/(v^2)}{U} + \frac{\alpha/(v\nu^2)}{U}\right)^2}{W}\right\}\\
\propto & \hspace{2mm}\pi(v)\frac{1}{v}U^{-\frac{1}{2}}W^{-\frac{1}{2}}\exp\left\{\frac{1}{2}\left[\frac{\alpha^2/(\nu^4)}{U} + \frac{\left(\frac{n_0\bar{y}_0}{\sigma_0^2} + \frac{\alpha/(v\nu^2)}{U}\right)^2}{W}\right]\right\} \times \\
&\exp\left\{\frac{n\bar{y}\theta}{\sigma^2}-\frac{n\theta^2}{2\sigma^2}-\frac{\theta^2}{2v}+\frac{1}{2}\frac{\theta^2/(v^2)}{U}+\frac{\theta\alpha/(v\nu^2)}{U}+\frac{1}{2}\frac{\left(\frac{\theta/(v^2)}{U}\right)^2+2\frac{\theta/(v^2)}{U}\left(\frac{n_0\bar{y}_0}{\sigma_0^2}+\frac{\alpha/(v\nu^2)}{U}\right)}{W}\right\}.
\end{align*}
\end{tiny}
Rewriting the part that involves $\theta$, we get 
\begin{align*}
&\exp\left\{\frac{n\bar{y}\theta}{\sigma^2}-\frac{n\theta^2}{2\sigma^2}-\frac{\theta^2}{2v}+\frac{1}{2}\frac{\theta^2/(v^2)}{U}+\frac{\theta\alpha/(v\nu^2)}{U}+\frac{1}{2}\frac{\left(\frac{\theta/(v^2)}{U}\right)^2+2\frac{\theta/(v^2)}{U}\left(\frac{n_0\bar{y}_0}{\sigma_0^2}+\frac{\alpha/(v\nu^2)}{U}\right)}{W}\right\},\\
&=\exp\left[-\frac{1}{2}(A\theta^2-2B\theta)\right],
\end{align*}
where $A=\frac{n}{\sigma^2}+\frac{1}{v}+\frac{1}{v^2U}-\frac{1}{(v^2U)^2W}$ and $B=\frac{n\bar{y}}{\sigma^2}+\frac{\alpha}{v\nu^2U} + \frac{\frac{n_0\bar{y}_0}{\sigma^2_0}+\frac{\alpha}{v\nu^2U}}{v^2UW}$.\\
When $\alpha=0$ and $\nu \rightarrow \infty$, $A\rightarrow\frac{n}{\sigma^2}+\frac{\frac{n_0}{\sigma_0^2}}{2v(\frac{n_0}{\sigma_0^2}+\frac{1}{2v})}$ and $B \rightarrow \frac{n\bar{y}}{\sigma^2} + \frac{n_0\bar{y}_0}{\sigma_0^2(1+\frac{2vn_0}{\sigma_0^2})}$. \\
We can see that, we obtain $\frac{B}{A}=\mu_p$ and $A^{-1}=\sigma^2_p$ if and only if $a_0=\frac{1}{1+\frac{2vn_0}{\sigma_0^2}}$. \\
Then $A=\frac{n}{\sigma^2}+\frac{n_0a_0}{\sigma_0^2}$ and $B = \frac{n\bar{y}}{\sigma^2} + \frac{n_0\bar{y}_0a_0}{\sigma_0^2}$. In addition, when $\alpha=0$ and $\nu \rightarrow \infty$, 
\begin{align*}
&\pi(v)\frac{1}{v}U^{-\frac{1}{2}}W^{-\frac{1}{2}}\exp\left\{\frac{1}{2}\left[\frac{\alpha^2/(\nu^4)}{U} + \frac{\left(\frac{n_0\bar{y}_0}{\sigma_0^2} + \frac{\alpha/(v\nu^2)}{U}\right)^2}{W}\right]\right\}\\
\rightarrow & \hspace{2mm}\pi(v)2v^{-\frac{1}{2}}\left(\frac{n_0}{\sigma^2_0}+\frac{1}{2v}\right)^{-\frac{1}{2}}\exp\left\{\frac{\frac{1}{2}(\frac{n_0\bar{y}_0}{\sigma_0^2})^2}{\frac{n_0}{\sigma^2_0}+\frac{1}{2v}}\right\},\\
\propto &\hspace{2mm}\pi(v)a_0^{\frac{1}{2}}\exp\left\{va_0\left(\frac{n_0\bar{y}_0}{\sigma_0^2}\right)^2\right\}.
\end{align*}
Then for the BHM, we have 
\begin{align*}
\pi(\theta, v|y, y_0) \propto & \hspace{2mm}\pi(v)a_0^{\frac{1}{2}}\exp\left\{va_0\left(\frac{n_0\bar{y}_0}{\sigma_0^2}\right)^2\right\}\exp\left[-\frac{1}{2}(A\theta^2-2B\theta)\right],\\
=&\hspace{2mm}\pi(v)a_0^{\frac{1}{2}}\exp\left\{va_0\left(\frac{n_0\bar{y}_0}{\sigma_0^2}\right)^2\right\}N\left(\frac{B}{A}, A^{-1}\right)A^{-\frac{1}{2}}\exp\left\{\frac{B^2}{2A}\right\},\\
=&\hspace{2mm}\pi(v)a_0^{\frac{1}{2}}\exp\left\{va_0\left(\frac{n_0\bar{y}_0}{\sigma_0^2}\right)^2\right\}N\left(\mu_p, \sigma^2_p\right)\sigma_p\exp\left\{\frac{\mu_p^2}{2\sigma_p^2}\right\}.
\end{align*}
For the NPP, we have 
\begin{align*}
\pi(\theta, a_0|y, y_0) \propto & \hspace{2mm}\pi(a_0)a_0^{\frac{1}{2}}\exp\left\{-\frac{n_0a_0}{2\sigma_0^2}\bar{y_{0}}^2\right\}N(\mu_p, \sigma_p^2)\sigma_p\exp\left\{\frac{\mu_p^2}{2\sigma_p^2}\right\}.
\end{align*}
Then a solution to 
\begin{equation}
    \label{seq:condition}
    \int\pi_{\text{NPP}}(\theta|f(v))\pi(f(v))\left\vert\frac{df(v)}{dv}\right\vert dv=\int\pi_{\text{BHM}}(\theta|v)\pi(v)dv
\end{equation}
is 
\begin{align*}
&\exp\left\{-\frac{n_0a_0}{2\sigma_0^2}\bar{y_{0}}^2\right\}\left\vert\frac{df(v)}{dv}\right\vert\pi(f(v))=\exp\left\{va_0\left(\frac{n_0\bar{y}_0}{\sigma_0^2}\right)^2\right\}\pi(v),\\
\Rightarrow & \exp\left\{-\frac{n_0a_0}{2\sigma_0^2}\bar{y_{0}}^2\left(1+2v\frac{n_0}{\sigma_0^2}\right)\right\}\left\vert\frac{df(v)}{dv}\right\vert\pi(f(v))=\pi(v),\\
\Rightarrow & \exp\left\{-\frac{n_0}{2\sigma_0^2}\bar{y_{0}}^2\right\}\left\vert\frac{df(v)}{dv}\right\vert\pi(f(v))=\pi(v).
\end{align*}
Since $\exp\left\{-\frac{n_0}{2\sigma_0^2}\bar{y_{0}}^2\right\}\left\vert\frac{df(v)}{dv}\right\vert \propto \left\vert\frac{df(v)}{dv}\right\vert$, then $$\frac{2n_0}{\sigma_0^2}\left(\frac{2vn_0}{\sigma_0^2}+1\right)^{-2}\pi(f(v))=\pi(v)$$ leads to a solution to (\ref{seq:condition}). 

\end{proof}

\subsection{Proof of Theorem \ref{multiple}}\label{appen:th2}
\begin{proof}

For the ANPP, we assume $a_{0k}=h_k(a_0)=c_{k}a_{0}$ and attempt to find the $c_k$ that will satisfy
\begin{equation}
    \label{seq:cond_2}
    \int\pi_{\text{ANPP}}(\theta|f(v))\pi(f(v))\left\vert\frac{df(v)}{dv}\right\vert dv=\int\pi_{\text{BHM}}(\theta|v)\pi(v)dv.
\end{equation}
The joint posterior of $\theta$ and $a_0$ is
$$\pi(\theta,a_0|y,y_0) \propto \exp\left\{-\frac{1}{2\sigma^2}\sum_{i=1}^n(y_i-\theta)^2\right\}\frac{\prod_{k=1}^{K}\left[\sigma_{0 k}^{-n_{0k}} \exp \left\{-\frac{1}{2 \sigma_{0 k}^{2}} \sum_{i=1}^{n_{0 k}}\left(y_{0ki}-\theta\right)^{2}\right\}\right]^{a_{0 k}} }{c(a_0)}\pi(a_0),$$
where $c(a_{0})=\int \prod_{k=1}^{K}\left[\sigma_{0 k}^{-n_{0k}} \exp \left\{-\frac{1}{2 \sigma_{0 k}^{2}} \sum_{i=1}^{n_{0 k}}\left(y_{0ki}-\theta\right)^{2}\right\}\right]^{a_{0 k}} d\theta$.

This then simplifies to
$$\exp\left\{-\frac{1}{2\sigma^2}\sum_{i=1}^n(y_i-\theta)^2\right\}\frac{ \exp \left\{-\sum_{k=1}^{K}\frac{a_{0k}}{2 \sigma_{0 k}^{2}} \sum_{i=1}^{n_{0 k}}\left(y_{0ki}-\theta\right)^{2}\right\}}{ \exp \left\{-\sum_{k=1}^{K}\frac{a_{0k}}{2 \sigma_{0 k}^{2}} (\sum_{i=1}^{n_{0 k}}y^2_{0ki}-n_{0k}M^2)\right\}\left[\sum_{k=1}^{K}\frac{n_{0k}a_{0k}}{2\pi\sigma^2_{0k}}\right]^{-\frac{1}{2}}}\pi(a_0),$$
where $M=\left(\sum_{k=1}^{K} \frac{a_{0k}}{\sigma_{0k}^{2}} \sum_{i=1}^{n_{0 k}} y_{0 k i}\right) /\left(\sum_{k=1}^{K} \frac{n_{0 k} a_{0 k}}{\sigma_{0 k}^{2}}\right)$.

Focusing on the power prior part, we get 
\begin{align*}
&\exp\left\{-\frac{1}{2\sigma^2}\sum_{i=1}^n(y_i-\theta)^2\right\}\exp \left\{-\sum_{k=1}^{K}\frac{c_ka_0}{2 \sigma_{0 k}^{2}} \sum_{i=1}^{n_{0 k}}\left(y_{0ki}-\theta\right)^{2}\right\},\\
\propto& \exp\left\{-\frac{1}{2}\left(\frac{n}{\sigma^2} + \sum_{k=1}^{K}\frac{c_kn_{0k}a_0}{\sigma^2_{0k}}\right)\left[\theta^2-2\mu_p\theta+\mu_p^2-\mu_p^2\right]\right\}\exp\left\{-\sum_{k=1}^{K}\frac{c_ka_0}{2\sigma^2_{0k}}\sum_{i=1}^{n_{0k}}y^2_{0ki}\right\},\\
=&\hspace{2mm}N(\mu_p, \sigma^2_p)\sigma_p\exp\left\{\frac{\mu^2_p}{2\sigma^2_p}\right\}\exp\left\{-\sum_{k=1}^{K}\frac{c_ka_0}{2\sigma^2_{0k}}\sum_{i=1}^{n_{0k}}y^2_{0ki}\right\},
\end{align*}
where $\sigma^2_p = \left(\frac{n}{\sigma^2} + \sum_{k=1}^{K}\frac{c_kn_{0k}a_0}{\sigma^2_{0k}}\right)^{-1}$ and $\mu_p = \sigma^2_p\left(\frac{n\bar{y}}{\sigma^2} + \sum_{k=1}^{K}\frac{c_kn_{0k}\bar{y}_{0k}a_0}{\sigma^2_{0k}}\right)$.\\
Further, $$c(a_0) = \exp \left\{-\sum_{k=1}^{K}\frac{a_{0k}}{2 \sigma_{0 k}^{2}} (\sum_{i=1}^{n_{0 k}}y^2_{0ki}-n_{0k}M^2)\right\}\left[\sum_{k=1}^{K}\frac{n_{0k}a_{0k}}{2\pi\sigma^2_{0k}}\right]^{-\frac{1}{2}},$$ where $M=\left(\sum_{k=1}^{K} \frac{a_{0k}}{\sigma_{0k}^{2}} \sum_{i=1}^{n_{0 k}} y_{0 k i}\right) /\left(\sum_{k=1}^{K} \frac{n_{0 k} a_{0 k}}{\sigma_{0 k}^{2}}\right)$.\\
We now move on to the BHM. 
The full posterior for the BHM is 
\begin{tiny}
\begin{align*}
\pi(\theta, \theta_0, \mu, v|y, y_0) \propto & \hspace{2mm}\pi(v)v^{-(K+1)/2} \times\\
&\exp\left\{\frac{-n(\bar{y}-\theta)^2}{2\sigma^2}-\frac{1}{2}\sum_{k=1}^{K}\frac{n_{0k}(\bar{y}_{0k}-\theta_{0k})^2}{\sigma^2_{0k}}-\frac{(\theta-\mu)^2}{2v}-\frac{1}{2v}\sum_{k=1}^{K}(\theta_{0k}-\mu)^2\right\}.
\end{align*}
\end{tiny}

Integrating out $\theta_0$, we get 
\begin{align*}
\pi(\theta, \theta_0, \mu, v|y, y_0) &\propto \pi(v)v^{-(K+1)/2}\exp\left\{\frac{-n(\bar{y}-\theta)^2}{2\sigma^2}-\frac{(\theta-\mu)^2}{2v}\right\} \times\\
&\int\exp\left\{-\frac{1}{2}\sum_{k=1}^{K}\frac{n_{0k}(\bar{y}_{0k}-\theta_{0k})^2}{\sigma^2_{0k}}-\frac{1}{2v}\sum_{k=1}^{K}(\theta_{0k}-\mu)^2\right\}d\theta_0.
\end{align*}
The integral above is
\begin{tiny}
\begin{align*}
&\int\exp\left\{-\frac{1}{2}\sum_{k=1}^{K}\frac{n_{0k}(\bar{y}_{0k}-\theta_{0k})^2}{\sigma^2_{0k}}-\frac{1}{2v}\sum_{k=1}^{K}(\theta_{0k}-\mu)^2\right\}d\theta_0,\\
\propto &\int\exp\left\{-\frac{1}{2}\left(\frac{\sum_{k=1}^{K}n_{0k}y^2_{0k}}{\sigma^2_{0k}}-2\sum_{k=1}^{K}\frac{n_{0k}\bar{y}_{0k}\theta_{0k}}{\sigma^2_{0k}}+\sum_{k=1}^{K}\frac{n_{0k}\theta^2_{0k}}{\sigma^2_{0k}}\right)-\frac{1}{2v}(\sum_{k=1}^{K}\theta^2_{0k}-2\mu\sum_{k=1}^{K}\theta_{0k}+K\mu^2) \right\}d\theta_0,\\
\propto &\int\exp\left\{-\frac{1}{2}\left(\sum_{k=1}^{K}\left(\frac{n_{0k}}{\sigma^2_{0k}}+\frac{1}{v}\right)\theta^2_{0k}-2\left[\sum_{k=1}^{K}\left(\frac{n_{0k}\bar{y}_{0k}}{\sigma^2_{0k}}+\frac{\mu}{v}\right)\theta_{0k}\right]\right) \right\}d\theta_0\exp\left\{-\frac{1}{2v}K\mu^2\right\},\\
=&\hspace{2mm}\exp\left\{-\frac{1}{2v}K\mu^2\right\}\exp\left\{\frac{1}{2}\sum_{k=1}^{K}\frac{(\frac{n_{0k}\bar{y}_{0k}}{\sigma^2_{0k}}+\frac{\mu}{v})^2}{\frac{n_{0k}}{\sigma^2_{0k}}+\frac{1}{v}}\right\}\prod_{k=1}^{K}\left(\frac{n_{0k}}{\sigma^2_{0k}}+\frac{1}{v}\right)^{-\frac{1}{2}}.
\end{align*}
\end{tiny}

Collecting all the terms involving $\theta_0$, $\mu$ and $v$, we get\\
\begin{tiny}
\begin{align*}
\pi(\theta, \mu, v|y_0, y) \propto & \hspace{2mm}\pi(v)v^{-(K+1)/2}\exp\left\{\frac{-n(\bar{y}-\theta)^2}{2\sigma^2}-\frac{(\theta-\mu)^2}{2v}-\frac{K\mu^2}{2v} + \frac{1}{2}\sum_{k=1}^{K}\frac{(\frac{n_{0k}\bar{y}_{0k}}{\sigma^2_{0k}}+\frac{\mu}{v})^2}{\frac{n_{0k}}{\sigma^2_{0k}}+\frac{1}{v}}\right\}\times\\
&\prod_{k=1}^{K}\left(\frac{n_{0k}}{\sigma^2_{0k}}+\frac{1}{v}\right)^{-\frac{1}{2}}.
\end{align*}
\end{tiny}
Integrating out $\mu$, we get 
\begin{align*}
\pi(\theta, v|y_0, y) \propto & \pi(v)v^{-(K+1)/2}\prod_{k=1}^{K}\left(\frac{n_{0k}}{\sigma^2_{0k}}+\frac{1}{v}\right)^{-\frac{1}{2}}\exp\left\{\frac{-n(\bar{y}-\theta)^2}{2\sigma^2}\right\}\\
&\int\exp\left\{-\frac{(\theta-\mu)^2}{2v}-\frac{K\mu^2}{2v} + \frac{1}{2}\sum_{k=1}^{K}\frac{(\frac{n_{0k}\bar{y}_{0k}}{\sigma^2_{0k}}+\frac{\mu}{v})^2}{\frac{n_{0k}}{\sigma^2_{0k}}+\frac{1}{v}}\right\}d\mu.
\end{align*}
The integral above is given by
\begin{align*}
&\int\exp\left\{-\frac{1}{2}\left[\left(\frac{1+K}{v} - \frac{1}{v^2}\sum_{k=1}^{K}\frac{1}{\frac{n_{0k}}{\sigma^2_{0k}}+\frac{1}{v}}\right)\mu^2 - 2\left(\frac{\theta}{v} + \sum_{k=1}^{K}\frac{\frac{n_{0k}\bar{y}_{0k}}{\sigma^2_{0k}v}}{\frac{n_{0k}}{\sigma^2_{0k}}+\frac{1}{v}}\right)\mu\right]\right\}d\mu\\
& \times \exp\left\{-\frac{1}{2}\left[\frac{\theta^2}{v}-\sum_{k=1}^{K}\frac{(\frac{n_{0k}\bar{y}_{0k}}{\sigma^2_{0k}})^2}{\frac{n_{0k}}{\sigma^2_{0k}}+\frac{1}{v}}\right]\right\},\\
&=\exp\left\{\frac{B^2}{2A}\right\}A^{-\frac{1}{2}}\exp\left\{-\frac{1}{2}\left[\frac{\theta^2}{v}-\sum_{k=1}^{K}\frac{(\frac{n_{0k}\bar{y}_{0k}}{\sigma^2_{0k}})^2}{\frac{n_{0k}}{\sigma^2_{0k}}+\frac{1}{v}}\right]\right\},
\end{align*}
where $A=\frac{1+K}{v} - \frac{1}{v^2}\sum_{k=1}^{K}\frac{1}{\frac{n_{0k}}{\sigma^2_{0k}}+\frac{1}{v}}$ and $B=\frac{\theta}{v} + \sum_{k=1}^{K}\frac{\frac{n_{0k}\bar{y}_{0k}}{\sigma^2_{0k}v}}{\frac{n_{0k}}{\sigma^2_{0k}}+\frac{1}{v}}$.\\
Collecting all terms involving $\theta$ and $v$, 
\begin{align*}
\pi(\theta, v|y_0, y) \propto & \hspace{2mm}\pi(v)v^{-(K+1)/2}\prod_{k=1}^{K}\left(\frac{n_{0k}}{\sigma^2_{0k}}+\frac{1}{v}\right)^{-\frac{1}{2}}A^{-\frac{1}{2}}\exp\left\{\frac{-n(\bar{y}-\theta)^2}{2\sigma^2}-\frac{\theta^2}{2v}+\frac{B^2}{2A}\right\}\\
&\times \exp\left\{\frac{1}{2}\left[\sum_{k=1}^{K}\frac{(\frac{n_{0k}\bar{y}_{0k}}{\sigma^2_{0k}})^2}{\frac{n_{0k}}{\sigma^2_{0k}}+\frac{1}{v}}\right]\right\}.\\
\end{align*}
Taking just the terms involving $\theta$, we get 
\begin{align*}
&\exp\left\{\frac{-n(\bar{y}-\theta)^2}{2\sigma^2}-\frac{\theta^2}{2v}+\frac{B^2}{2A}\right\},\\
\propto & \exp\left\{-\frac{1}{2}\left[\left(\frac{n}{\sigma^2} + \frac{1}{v} - \frac{1}{v^2A}\right)\theta^2 -2\left(\frac{n\bar{y}}{\sigma^2} + \sum_{k=1}^{K}\frac{\frac{n_{0k}\bar{y}_{0k}}{\sigma^2_{0k}v^2}}{(\frac{n_{0k}}{\sigma^2_{0k}}+\frac{1}{v})A}\right)\theta - \frac{\left(\sum_{k=1}^{K}\frac{\frac{n_{0k}\bar{y}_{0k}}{\sigma^2_{0k}}}{v(\frac{n_{0k}}{\sigma^2_{0k}}+\frac{1}{v})}\right)^2}{A}\right]\right\}.
\end{align*}

Then the full posterior is 
\begin{tiny}
\begin{align*}
&\pi(\theta, v|y, y_0)\\
\propto &\hspace{2mm}\pi(v)v^{-(K+1)/2}\prod_{k=1}^{K}\left(\frac{n_{0k}}{\sigma^2_{0k}}+\frac{1}{v}\right)^{-\frac{1}{2}}A^{-\frac{1}{2}}\exp\left\{\frac{\left(\sum_{k=1}^{K}\frac{\frac{n_{0k}\bar{y}_{0k}}{\sigma^2_{0k}}}{v(\frac{n_{0k}}{\sigma^2_{0k}}+\frac{1}{v})}\right)^2}{2A}\right\}\exp\left\{\frac{1}{2}\sum_{k=1}^{K}\frac{(\frac{n_{0k}\bar{y}_{0k}}{\sigma^2_{0k}})^2}{\frac{n_{0k}}{\sigma^2_{0k}}+\frac{1}{v}}\right\}\\
&\times\sigma_{b}N(\mu_{b}, \sigma^2_{b})\exp\left\{\frac{\mu^2_{b}}{2\sigma^2_{b}}\right\},
\end{align*}
\end{tiny}
where $\sigma^2_{b}=\left(\frac{n}{\sigma^2} + \frac{1}{v} - \frac{1}{v^2A}\right)^{-1}$ and $\mu_{b} = \left(\frac{n\bar{y}}{\sigma^2} + \sum_{k=1}^{K}\frac{\frac{n_{0k}\bar{y}_{0k}}{\sigma^2_{0k}v^2}}{(\frac{n_{0k}}{\sigma^2_{0k}}+\frac{1}{v})A}\right)\sigma^2_{b}$.\\
We need to find $a_0$ and $c_k$ in terms of $v$ that can fulfill $\sigma^2_{b}=\sigma^2_p$ and $\mu_{b}=\mu_p$.
We first set $\sigma^2_p=\sigma^2_{b}$ to get
$$\left(\frac{n}{\sigma^2} + \frac{1}{v} - \frac{1}{v^2A}\right)^{-1}=\left(\frac{n}{\sigma^2} + \sum_{k=1}^{K}\frac{c_kn_{0k}a_0}{\sigma^2_{0k}}\right)^{-1},$$
which gives 
$$a_0 = \frac{v^{-1}-v^{-2}A^{-1}}{\sum_{k=1}^{K}\frac{c_kn_{0k}}{\sigma^2_{0k}}} = \frac{v^{-1}-v^{-2}\left(\frac{K+1}{v}-\sum\limits_{k=1}^{K}\frac{1}{v^2\left(\frac{n_{0k}}{\sigma^2_{0k}}+\frac{1}{v}\right)}\right)^{-1}}{\sum_{k=1}^{K}\frac{c_kn_{0k}}{\sigma^2_{0k}}}.$$
We then set $\mu_p = \mu_{b}$ and get
$$\sigma^2_p\left(\frac{n\bar{y}}{\sigma^2} + \sum_{k=1}^{K}\frac{c_kn_{0k}\bar{y}_{0k}a_0}{\sigma^2_{0k}}\right)=\left(\frac{n\bar{y}}{\sigma^2} + \sum_{k=1}^{K}\frac{\frac{n_{0k}\bar{y}_{0k}}{\sigma^2_{0k}v^2}}{(\frac{n_{0k}}{\sigma^2_{0k}}+\frac{1}{v})A}\right)\sigma^2_{b}.$$
Since $$v^{-1}-v^{-2}A^{-1}=\frac{vK-\sum_{k=1}^{K}N_k}{v\left[v(1+K)-\sum_{k=1}^{K}N_k\right]},$$
where $N_k=\left(\frac{n_{0k}}{\sigma^2_{0k}}+\frac{1}{v}\right)^{-1}$, $$\sum_{k=1}^{K}\frac{\frac{n_{0k}\bar{y}_{0k}}{\sigma^2_{0k}v^2}}{(\frac{n_{0k}}{\sigma^2_{0k}}+\frac{1}{v})A}=\sum_{k=1}^{K}\frac{n_{0k}\bar{y}_{0k}}{\sigma^2_{0k}}\frac{N_k}{\left[v(1+K)-\sum_{k=1}^{K}N_k\right]},$$
and $$\sum_{k=1}^{K}\frac{c_kn_{0k}\bar{y}_{0k}a_0}{\sigma^2_{0k}}=\frac{\sum_{k=1}^{K}c_kn_{0k}\bar{y}_{0k}/\sigma^2_{0k}}{\sum_{k=1}^{K}c_kn_{0k}/\sigma^2_{0k}}\frac{K-\sum_{k=1}^{K}N_k/v}{\left[v(1+K)-\sum_{k=1}^{K}N_k\right]},$$
the expression simplifies to
$$\sum_{k=1}^{K}\frac{n_{0k}\bar{y}_{0k}}{\sigma^2_{0k}}N_k=\frac{\sum_{k=1}^{K}c_kn_{0k}\bar{y}_{0k}/\sigma^2_{0k}}{\sum_{k=1}^{K}c_kn_{0k}/\sigma^2_{0k}}\left(K-\sum_{k=1}^{K}N_k/v\right).$$ If we put $$c_k = N_k/v=\frac{1}{1+\frac{n_{0k}v}{\sigma^2_{0k}}},$$ then $\left(K-\sum_{k=1}^{K}N_k/v\right)=v\left(\sum_{k=1}^{K}c_kn_{0k}/\sigma^2_{0k}\right)$, and (\ref{seq:cond_2}) is satisfied.
\\ Plugging in $c_k=\left(1+\frac{n_{0k}v}{\sigma^2_{0k}}\right)^{-1}$ into the expression for $a_0$, we get 
\begin{align*}
a_0=&\left(\sum_{k=1}^{K}\frac{N_kn_{0k}}{v\sigma^2_{0k}}\right)^{-1}\frac{K-\sum_{k=1}^{K}N_k/(v)}{v(1+K)-\sum_{k=1}^{K}N_k},\\
&=\frac{v}{v(1+K)-\sum_{k=1}^{K}N_k},\\
&=\frac{1}{(1+K)-\sum_{k=1}^{K}\left(1+\frac{n_{0k}v}{\sigma^2_{0k}}\right)^{-1}},\\
&=\left[1+\sum_{k=1}^{K}\frac{\frac{n_{0k}}{\sigma^2_{0k}}}{\frac{n_{0k}}{\sigma^2_{0k}}+\frac{1}{v}}\right]^{-1},
\end{align*}
and
\begin{align*}
c_ka_0 &=\left(1+\frac{n_{0k}v}{\sigma^2_{0k}}\right)^{-1}\left[1+\sum_{k=1}^{K}\frac{\frac{n_{0k}}{\sigma^2_{0k}}}{\frac{n_{0k}}{\sigma^2_{0k}}+\frac{1}{v}}\right]^{-1}\\
&=\left[1+\frac{n_{0k}v}{\sigma^2_{0k}}+\left(1+\frac{n_{0k}v}{\sigma^2_{0k}}\right)\sum_{k=1}^{K}\frac{\frac{n_{0k}}{\sigma^2_{0k}}}{\frac{n_{0k}}{\sigma^2_{0k}}+\frac{1}{v}}\right]^{-1}.
\end{align*}
We can see that after the transformations, we get $0 < a_0 < 1$ and $0 < c_ka_0 < 1$ as desired.\\
Then we further simplify the joint posterior of $\theta$ and $a_0$ with a ANPP:
\begin{tiny}
\begin{align*}
&\pi_{\text{ANPP}}(\theta,a_0|y,y_0)\\
\propto&\hspace{2mm}N(\mu_p, \sigma^2_p)\sigma_p\exp\left\{\frac{\mu^2_p}{2\sigma^2_p}\right\}\exp\left\{-\sum_{k=1}^{K}\frac{c_ka_0}{2\sigma^2_{0k}}\sum_{i=1}^{n_{0k}}y^2_{0ki}\right\}\exp \left\{\sum_{k=1}^{K}\frac{a_{0k}}{2 \sigma_{0 k}^{2}} (\sum_{i=1}^{n_{0 k}}y^2_{0ki}-n_{0k}M^2)\right\}\\
& \times \left[\sum_{k=1}^{K}\frac{n_{0k}a_{0k}}{2\pi\sigma^2_{0k}}\right]^{\frac{1}{2}}\pi(a_0),\\
\propto& \hspace{2mm}N(\mu_p, \sigma^2_p)\sigma_p\exp\left\{\frac{\mu^2_p}{2\sigma^2_p}\right\}\exp\left\{-\sum_{k=1}^{K}\frac{a_{0k}n_{0k}}{2\sigma_{0k}^{2}}\left(\frac{\sum_{k=1}^{K} \frac{a_{0k}}{\sigma_{0k}^{2}} \sum_{i=1}^{n_{0 k}} y_{0 k i}}{\sum_{k=1}^{K} \frac{n_{0 k} a_{0 k}}{\sigma_{0 k}^{2}}}\right)^2\right\}\left[\sum_{k=1}^{K}\frac{n_{0k}a_{0k}}{2\pi\sigma^2_{0k}}\right]^{\frac{1}{2}}\pi(a_0),\\
\propto& \hspace{2mm}N(\mu_p, \sigma^2_p)\sigma_p\exp\left\{\frac{\mu^2_p}{2\sigma^2_p}\right\}\exp\left\{-\frac{\left(\sum_{k=1}^{K} \frac{n_{0k}a_{0k}\bar{y}_{0k}}{\sigma_{0k}^{2}}\right)^2}{2\sum_{k=1}^{K} \frac{n_{0k} a_{0k}}{\sigma_{0k}^{2}}}\right\}\left[\sum_{k=1}^{K}\frac{n_{0k}a_{0k}}{2\pi\sigma^2_{0k}}\right]^{\frac{1}{2}}\pi(a_0),\\
\propto& \hspace{2mm}N(\mu_p, \sigma^2_p)\sigma_p\exp\left\{\frac{\mu^2_p}{2\sigma^2_p}\right\}\exp\left\{-\frac{\left(\sum_{k=1}^{K}\frac{n_{0k}c_ka_0\bar{y}_{0k}}{\sigma_{0k}^{2}}\right)^2}{2\sum_{k=1}^{L_{0}} \frac{n_{0k}c_ka_0}{\sigma_{0k}^{2}}}\right\}\left[\sum_{k=1}^{K}\frac{n_{0k}c_ka_0}{2\pi\sigma^2_{0k}}\right]^{\frac{1}{2}}\pi(a_0),\\
\propto& \hspace{2mm}N(\mu_p, \sigma^2_p)\sigma_p\exp\left\{\frac{\mu^2_p}{2\sigma^2_p}\right\}\exp\left\{-\frac{a_0}{2C}\left(\sum_{k=1}^{K}c_kY_k\right)^2\right\}\left[a_0C\right]^{\frac{1}{2}}\pi(a_0),
\end{align*}
\end{tiny}
where $C=\sum_{k=1}^{K}\frac{n_{0k}c_k}{\sigma^2_{0k}}$ and $Y_k=\frac{n_{0k}\bar{y}_{0k}}{\sigma^2_{0k}}$. 

The joint posterior of $\theta$ and $v$ with a BHM is
\begin{align*}
&\pi(\theta, v|y, y_0)\\
\propto &\hspace{2mm}\pi(v)v^{-(K+1)/2}\prod_{k=1}^{K}N_k^{\frac{1}{2}}A^{-\frac{1}{2}}\exp\left\{\frac{\left(\sum_{k=1}^{K}Y_kN_k\right)^2}{2v^2A}\right\}\exp\left\{\frac{1}{2}\sum_{k=1}^{K}Y_k^2N_k\right\}\\
&\sigma_{b}N(\mu_{b}, \sigma^2_{b})\exp\left\{\frac{\mu^2_{b}}{2\sigma^2_{b}}\right\}.
\end{align*}
Since $a_0=f(v)=\left[1+\sum_{k=1}^{K}\frac{\frac{n_{0k}}{\sigma^2_{0k}}}{\frac{n_{0k}}{\sigma^2_{0k}}+\frac{1}{v}}\right]^{-1}$, the Jacobian is 
\begin{align*}
\left|\frac{df(v)}{dv}\right|=-\left[1+\sum_{k=1}^{K}\frac{\frac{n_{0k}}{\sigma^2_{0k}}}{\frac{n_{0k}}{\sigma^2_{0k}}+\frac{1}{v}}\right]^{-2}\left[\sum_{k=1}^{K}\frac{n_{0k}}{\sigma^2_{0k}}(\frac{n_{0k}}{\sigma^2_{0k}}+\frac{1}{v})^{-2}v^{-2}\right].
\end{align*}
One solution to the equation $$\int\pi_{\text{ANPP}}(\theta|f(v))\pi(f(v))\left\vert\frac{df(v)}{dv}\right\vert dv=\int\pi_{\text{BHM}}(\theta|v)\pi(v)dv$$ is 
$$\pi_{\text{ANPP}}(\theta, f(v))\left|\frac{df(v)}{dv}\right|=\pi_{BHM}(\theta, v).$$
Since we have $\sigma_b=\sigma_p$ and $\mu_b=\mu_p$, and denote
$$Q(a_0)=\exp\left\{-\frac{a_0}{2C}\left(\sum_{k=1}^{K}c_kY_k\right)^2\right\}\left[a_0C\right]^{\frac{1}{2}}$$ and 
$$R(v)=v^{-(K+1)/2}\prod_{k=1}^{K}N_k^{\frac{1}{2}}A^{-\frac{1}{2}}\exp\left\{\frac{\left(\sum_{k=1}^{K}Y_kN_k\right)^2}{2v^2A}\right\}\exp\left\{\frac{1}{2}\sum_{k=1}^{K}Y_k^2N_k\right\},$$
then equation \eqref{seq:cond_2} is satisfied if $$Q(f(v)) \cdot \left\vert\frac{df(v)}{dv}\right\vert \pi(f(v)) = R(v) \cdot \pi(v).$$
\end{proof}

\section{Additional Figures}\label{appen:fig}

\begin{figure}[H]
\centering
\includegraphics[width=0.9\textwidth]{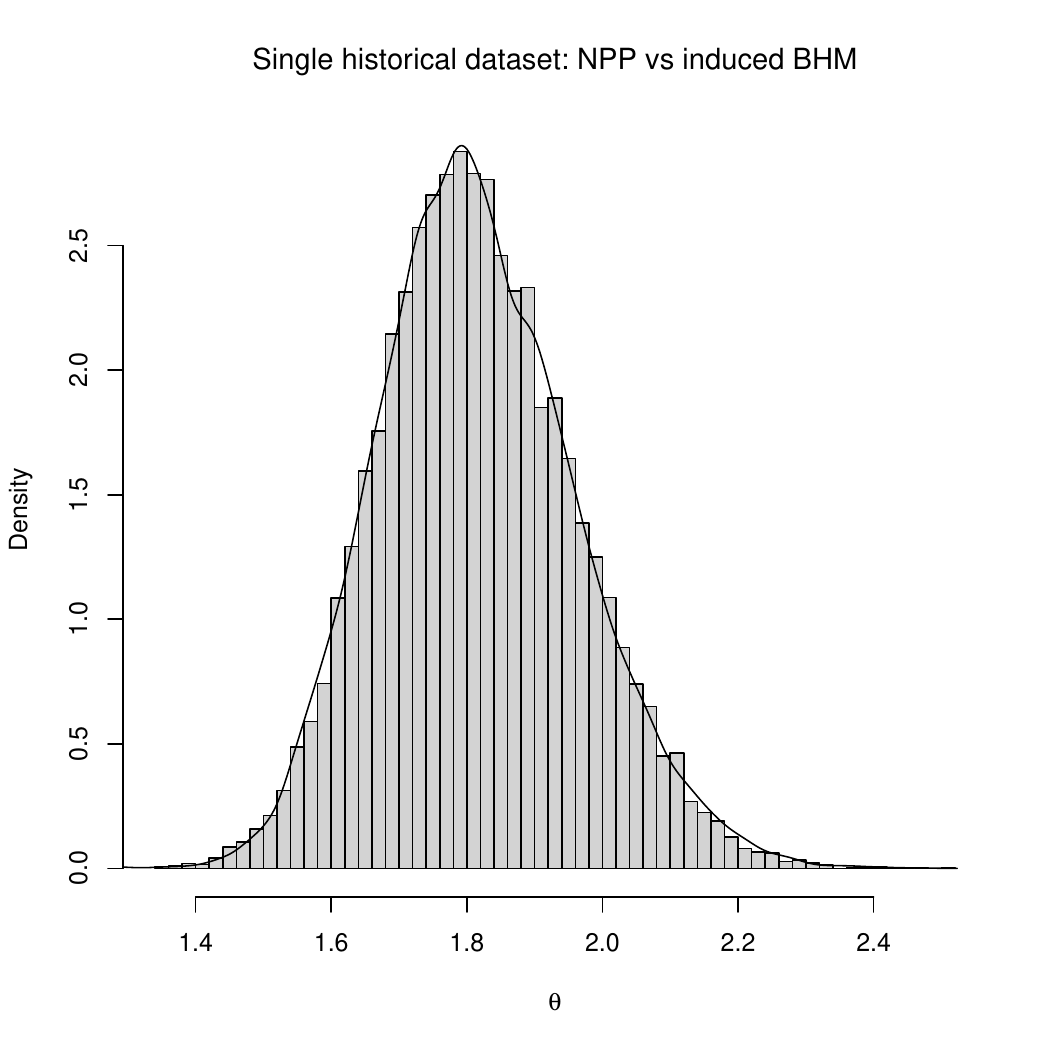}
\captionsetup{width=0.8\linewidth}
\caption{Single historical dataset: in this example, $n=n_0=20$, $\bar{y}=2$, $\bar{y}_0=1.5$, $\sigma^2=0.5$ and $\sigma^2_0=0.3$. We ran four independent chains of 10,000 iterations with 5,000 burn-ins using RStan. The histogram represents the posterior of $\theta$ using a normalized power prior with a beta($2$, $2$) prior on $a_0$. The density curve represents the posterior of $\theta$ using the BHM where the prior on $v$ is induced using Theorem \ref{single}. We observe that the two posteriors are equivalent.}
\label{test_single}
\end{figure}

\begin{figure}[H]
\centering
\includegraphics[width=0.9\textwidth]{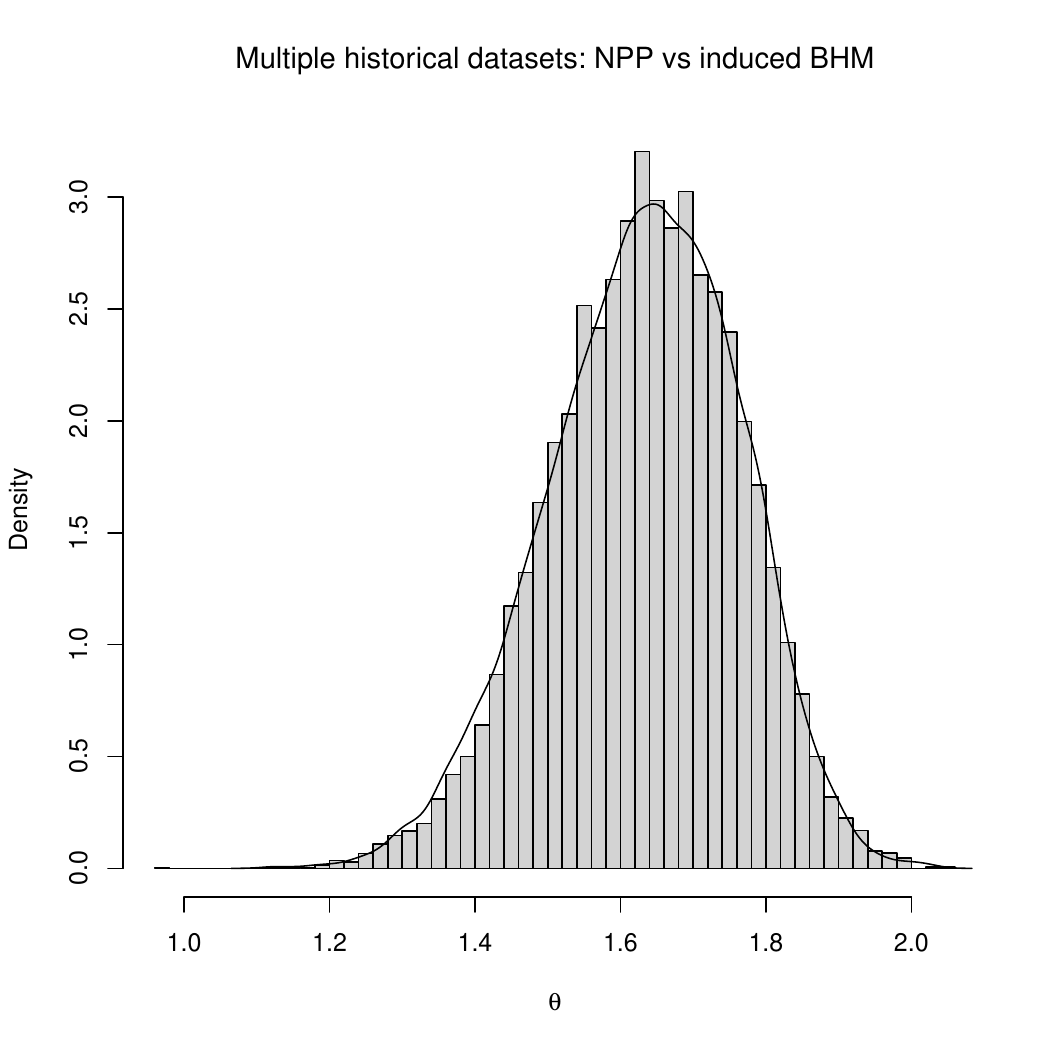}
\captionsetup{width=0.8\linewidth}
\caption{Multiple historical datasets: in this example, there are three historical datasets, and $n=30$, $n_0=(20,30,50)$, $\sigma^2=0.5$, $\sigma^2_0=(0.5,1,1.5)$, $\bar{y}_0=(1,2,3)$, and $\bar{y}=1.5$. We ran four independent chains of 8,000 iterations with 4,000 burn-ins using RStan. The histogram represents the posterior of $\theta$ using a normalized power prior with beta($2$, $2$) prior on $a_0$ and $h_k(a_0)$ chosen according to Theorem \ref{multiple}. The density curve represents the posterior of $\theta$ using the BHM where the prior on $v$ is induced using Theorem \ref{multiple}. We observe that the two posteriors are equivalent.}
\label{test_multiple}
\end{figure}

\begin{figure}[H]
\centering
\includegraphics[width=0.9\textwidth]{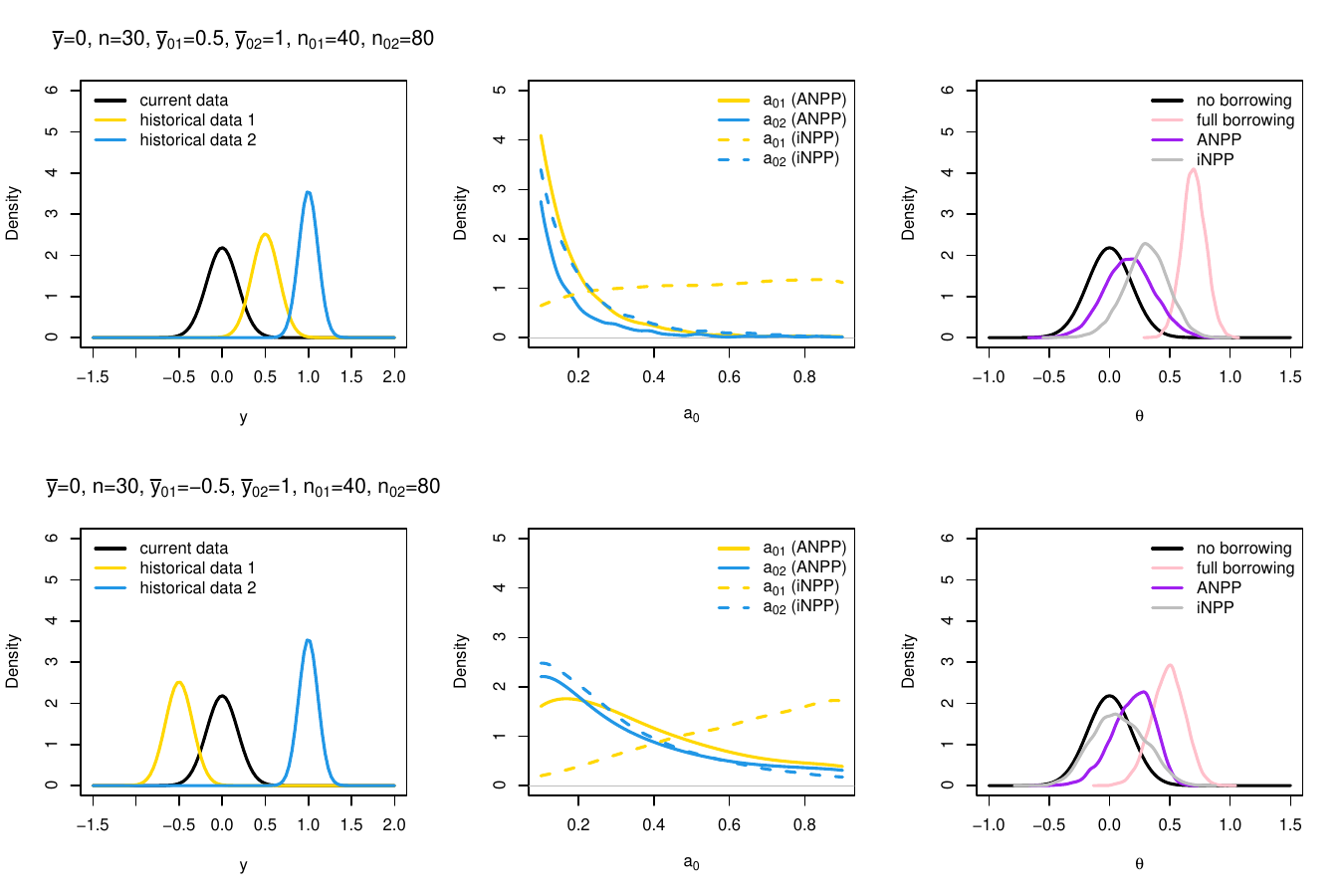}
\captionsetup{width=0.8\linewidth}
\caption{Marginal posteriors of $a_{01}$, $a_{02}$ and $\theta$ using the ANPP and the independent NPP for simulated i.i.d. normal data. In these scenarios, one of the historical datasets is incompatible with the current data while the other is partially compatible. The first column includes plots of the densities of the current data (black line) and two historical datasets (yellow and blue lines). The second column includes plots of the posterior densities of $a_{01}$ and $a_{02}$ using the ANPP (yellow and blue solid lines) and the iNPP (yellow and blue dashed lines). The third column includes plots of the posterior densities of $\theta$ using four different priors, the ANPP (purple), the iNPP (grey), the power prior with $a_0=0$ (black) and the power prior with $a_0=1$ (pink).}
\label{multiple_sim3}
\end{figure}

\begin{figure}[H]
\centering
\includegraphics[width=0.9\textwidth]{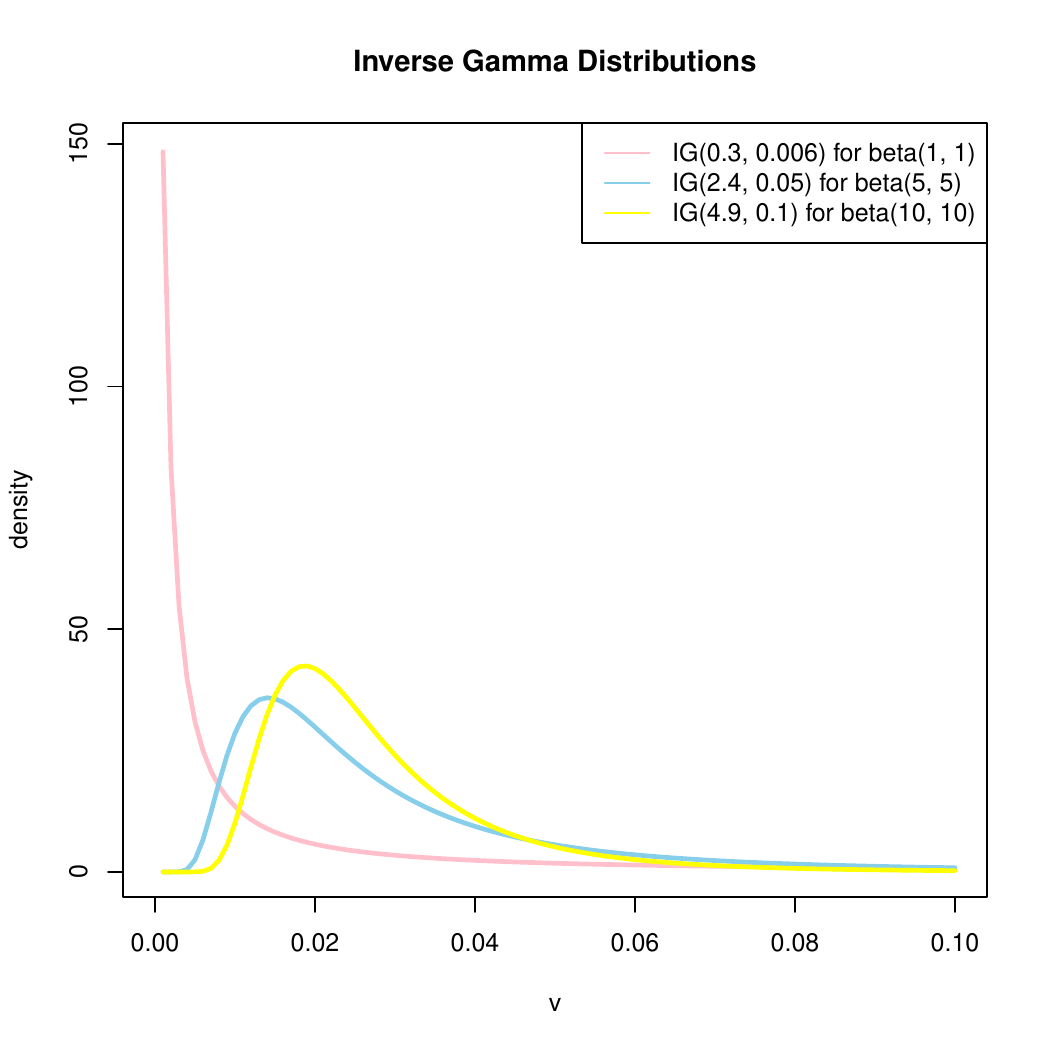}
\captionsetup{width=0.8\linewidth}
\caption{We choose three beta distributions with mean equal to $0.5$ as the prior for $a_0$ in the normalized power prior, and find the best approximating inverse gamma distribution prior for $v$ in a BHM. We observe that as we borrow more historical information, i.e. as the variance of the beta distribution decreases, the variance of the prior on $v$ also decreases, i.e. more historical information is borrowed.}
\label{IG}
\end{figure}

\end{document}